\makeatletter
\@ifundefined{@parse@version@dash}{%
\def\@parse@version#1{\@parse@version@0#1}
\def\@parse@version@#1/#2/#3#4#5\@nil{%
\@parse@version@dash#1-#2-#3#4\@nil}
\def\@parse@version@dash#1-#2-#3#4#5\@nil{%
  \if\relax#2\relax\else#1\fi#2#3#4 }
}{}
\makeatother
\documentclass[reprint,superscriptaddress, amsmath,amssymb, pra]{revtex4-2}
\usepackage{graphicx}
\usepackage{subfigure}
\usepackage{dcolumn}
\usepackage{bm}
\usepackage{color}
\usepackage{ulem}

\begin{document}
\preprint{APS/123-QED}

\title{Robust quantum teleportation via a non-Markovian channel}

\author{Yanfang Wang}
\author{Shibei Xue}%
\email{shbxue@sjtu.edu.cn}
\affiliation{
Department of Automation, Shanghai Jiao Tong University, Shanghai 200240, People's Republic of China}
\affiliation{Key Laboratory of System Control and Information Processing,
 Ministry of Education of China, Shanghai 200240, People's Republic of China}
\affiliation{Shanghai Engineering Research Center of Intelligent Control and Management, Shanghai 200240, People's Republic of China}%
\author{Hongbin Song}%
\email{songhongbin@cuhk.edu.cn}
\affiliation{General Education Division and Shenzhen Key Lab of Semiconductor Lasers, The Chinese University of Hong Kong, Shenzhen, Guang Dong 518116, People's Republic of China}%
\author{Min Jiang}
\email{jiangmin08@suda.edu.cn}
\affiliation{School of Electronics and Information Engineering, Soochow University, Suzhou 215006, People's Republic of China
}%

\begin{abstract}
In this paper, we propose a non-Markovian quantum channel approach to mitigating the degradation of the average fidelity in continuous-variable quantum teleportation. The non-Markovian quantum channel is modeled by an augmented system, where ancillary systems are introduced to represent the internal modes of non-Markovian environments. With a proper non-Markovianity, enhanced effects of the channels on continuous-variable quantum teleportation are observed. Also, the logarithmic negativity of entangled states can be effectively maintained so that the decay of the average fidelity of quantum teleportation is mitigated. In addition, the analysis on different teleported quantum states shows that coherent states exhibit stronger robustness than those of squeezed states and cat states in quantum teleportation with a non-Markovian channel.
\end{abstract}

\maketitle


\section{\label{sec:level1} introduction}
Quantum teleportation based on quantum entanglement can safely teleport quantum states carrying secret quantum information from one location to a distant one. It was firstly proposed and demonstrated based on discrete-variable Einstein-Podolsky-Rosen (EPR) pairs by Bennett and Bouwmeeste, respectively~\cite{bennett1993teleporting,bouwmeester1997experimental}. Based on quantum teleportation, various protocols and technologies have been developed, such as quantum gate teleportation~\cite{aliferis2004computation}, port-based teleportation~\cite{ishizaka2008asymptotic,ishizaka2009quantum}, and quantum repeaters~\cite{briegel1998quantum}, etc. 
Besides discrete-variable protocols, continuous-variable (CV) quantum teleportation was proposed by Vaidman~\cite{vaidman1994teleportation}, followed which teleportation of coherent states, squeezed states and cat states have been experimentally demonstrated~\cite{braunstein1998teleportation, furusawa1998unconditional, takei2005experimental,lee2011teleportation} to construct quantum communication networks and quantum computers~\cite{briegel1998quantum,fukui2022building}. However, the performance of a CV quantum teleportation system is determined by the quality of quantum entanglement~\cite{bennett1993teleporting}, which is significantly affected by the disturbance in a quantum channel.

 To counteract the unexpected impact of the channel, various protocols have been proposed to enhance the quantum entanglement on quantum sources. For example, in Ref.~\cite{hu2021long}, an entanglement purification approach was proposed, using only one pair of hyperentangled state. Ref.~\cite{riera2021entanglement} presented entanglement-assisted entanglement purification protocols by utilizing auxiliary, high-dimensional entangled states. Deterministic entanglement purification protocols and measurement-based entanglement purification protocols were introduced in Ref.~\cite{yan2023advances}.
 However, few work solves such problems from the perspective of quantum channels. Actually, optical fibers are widely employed as prominent transmission channels in quantum communication, where noises are attributed to Raman reservoirs and Brillouin scattering~\cite{drummond2001quantum}. Such noise can be modeled by Lorentzian noise~\cite{drummond2001quantum}. Lorentzian noise is a kind of quantum colored noise that can give rise to non-Markovian dynamics of quantum systems. Therefore, it is worthwhile to investigate the effects of a noisy channel that exhibits non-Markovian dynamics; i.e., parts of leaked information can be fed back to the transmitted state, and the performance of CV quantum teleportation would be improved.
In this paper, we investigate the effect of a non-Markovian channel on CV quantum teleportation. The non-Markovian channel is modeled as an augmented system, where ancillary systems are introduced to represent the internal modes of a non-Markovian channel. With this model, the dynamics of the transmitted state in the channel can be obtained from partial trace of time evolution of the density matrix of the augmented system. Together with the generation of entangled states and measurement processes, the performance of CV quantum teleportation can be evaluated by an average fidelity.  Moreover, how non-Markovianity affects the fidelity is also investigated, where we find a proper non-Markovianity can help to achieve a robust CV quantum teleportation, keeping the fidelity on a high level.
In addition, different states including coherent states, cat states and squeezed states are transmitted in the non-Markovian channels, whose performances are also compared in this paper.


This paper is organized as follows. In Section II, we describe CV quantum teleportation in the Schrodinger picture. Based on the augmented system, CV quantum teleportation with a non-Markovian quantum channel is presented in Section III. In Section IV, we analyze the effects of the non-Markovian dynamics of a quantum channel disturbed by Lorentzian noise and two-Lorentzian noise. Then the influence of different input states is investigated in Section V. Finally, we draw conclusions in Section VI.

\section{continuous variable quantum teleportation}
In CV quantum teleportation, information is encoded on continuous-variable quantum states and then reliably transferred to a remote receiver through a quantum entanglement channel assisted by a classical communication channel.
Here, we will briefly review CV quantum teleportation and its corresponding notations are introduced. After that, two relevant measures for evaluating the performance of CV quantum teleportation are introduced as well.

\subsection{Basic process of CV quantum teleportation}
\begin{figure}[ht]
\includegraphics[scale=0.7]{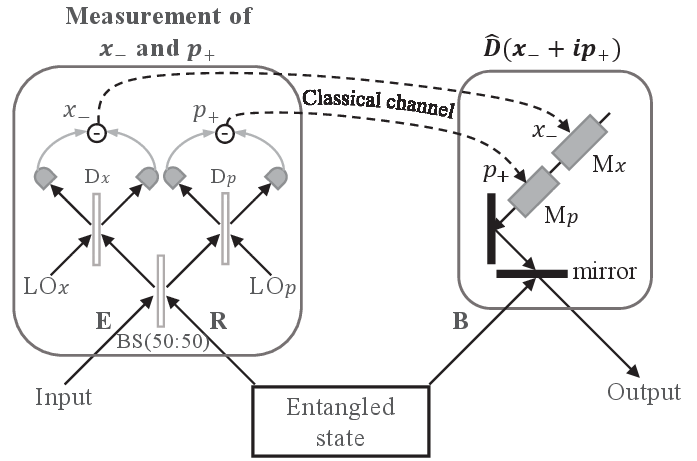}
\caption{\label{CV quantum teleportation}
Schematic representation of CV quantum teleportation~\cite{furusawa1998unconditional}. BS represents a beam splitter. $\mathrm{D}_x$ and $\mathrm{D}_p$ represent the balanced homodyne detectors for $x$ and $p$ quadrature components, respectively. $\mathrm{LO}_x$ and $\mathrm{LO}_p$ represent local oscillators (LOs) that are adjusted to $x$ and $p$ quadrature, respectively. $\mathrm{M}_x$ and $\mathrm{M}_p$ represent the amplitude and phase modulators, respectively. Mirror refers to a high reflectance mirror.
}
\end{figure}

A schematic diagram of CV quantum teleportation is shown in Fig.~\ref{CV quantum teleportation}. The sender Alice and the receiver Bob first share an entangled state $|q(\mathrm{R},\mathrm{B})\rangle$ which is commonly a two-mode squeezed vacuum state~\cite{walls1994atomic} expressed as
\begin{equation}
|q(\mathrm{R},\mathrm{B})\rangle = \sqrt{1-q^2} \sum_{n=0}^{\infty} q^n |n\rangle_{\mathrm{R}} \otimes |n\rangle_{\mathrm{B}},
\end{equation}
where $q=tanh(r)$ and $r$ is a squeezing parameter. Note that when $r \rightarrow \infty$, then $q \rightarrow 1$, the entangled squeezed state approaches the Einstein-Podolsky-Rosen (EPR) state.

Further, Alice conducts a Bell measurement by mixing the input state $|\psi_{\mathrm{E}}\rangle$ with mode $\mathrm{R}$ of the entangled state by a balanced beam splitter. Then, she performs measurements of $\hat{x}_{-}$ and  $\hat{p}_{+}$ by balanced homodyne detectors ($\mathrm{D}_x$, $\mathrm{D}_p$), where the local oscillators (LOs) of $\mathrm{D}_x$ and $\mathrm{D}_p$ are set to $x$ and $p$ quadratures, respectively~\cite{furusawa1998unconditional,furusawa2007quantum}. Subsequently,
Alice obtains the values of  $x_{-}$ and $p_{+}$,
where $\hat{x}_{-}= (\hat{x}_{\mathrm{E}}-\hat{x}_\mathrm{R})/\sqrt{2}$ , $\hat{p}_{+}=(\hat{p}_{\mathrm{E}}+\hat{p}_{\mathrm{R}})/\sqrt{2}$, in which $\hat{x}_\mathrm{E}$, $\hat{p}_\mathrm{E}$ and $\hat{x}_\mathrm{R}$, $\hat{p}_\mathrm{R}$ are the quadrature components of the input state and mode $\mathrm{R}$ of the entangled state, respectively. The eigenstates of $\hat{x}_{-}$ and $\hat{p}_{+}$ after measurement can be expressed as~\cite{hofmann2000fidelity}
 \begin{equation}
 |x_{-},p_{+}\rangle = \frac{1}{\sqrt{\pi}} \sum_{n=0}^\infty \hat{D}_{\mathrm{E}}(x_{-}+ip_{+})|n\rangle_{\mathrm{E}}\otimes |n\rangle_{\mathrm{R}},
 \label{measure_basis}
 \end{equation}
 where
 \begin{eqnarray}
 \hat{D}_{\mathrm{E}}(x_{-}+ip_{+})=&&\mathrm{exp}[2i(p_{+}\hat{x}_{\mathrm{E}}-x_{-}\hat{p}_{\mathrm{E}})] \nonumber\\
 \end{eqnarray}
 is a displacement operator.
Meanwhile, the measurement projects the initial state of the overall system described by $|\psi_{\mathrm{E}}\rangle \otimes |q(\mathrm{R},\mathrm{B})\rangle$ into the quantum state of mode $\mathrm{B}$~\cite{hofmann2000fidelity}
\begin{equation}
|\psi_\mathrm{B}(x_{-},p_{+})\rangle = \sqrt{\frac{1-q^2}{\pi}} \sum_{n=0}^\infty q^n |n\rangle \langle n| \hat{D}_{\mathrm{E}}(-x_{-}-ip_{+})|\psi_{\mathrm{E}}\rangle.
\end{equation}

Finally, through a classical communication channel, Bob gets the values of $x_{-}$ and $p_{+}$ from Alice and then displaces his state by the displacement operator $\hat{D}_{\mathrm{E}}(x_{-}+ip_{+})$. The displacement operation can be realized with two modulators and a high reflectance mirror, as shown in Fig.~\ref{CV quantum teleportation}. The output state is thus written as~\cite{furusawa2007quantum},
\begin{eqnarray}
|\psi_{out}(x_{-},p_{+})\rangle =&& \sqrt{\frac{1-q^2}{\pi}} \sum_{n=0}^\infty q^n \hat{D}_{\mathrm{E}}(x_{-}+ip_{+})|n\rangle \langle n| \nonumber\\
 &&\hat{D}_{\mathrm{E}}(-x_{-}-ip_{+})|\psi_{\mathrm{E}}\rangle.
\end{eqnarray}
When the entangled state approaches the EPR state; i.e., $q \to 1$, the output state approaches exactly the input state $|\psi_{\mathrm{E}}\rangle$. In this case, the information on the input state can be successfully reconstructed by Bob. Otherwise, the state reconstructed by Bob is not perfect such that the fidelity of the teleportation degrades, which will be introduced in the subsection.

\subsection{Figure of merit of CV quantum teleportation}
In CV quantum teleportation, the entangled state plays a critical role, and its entanglement directly determines the performance of information transmission.
In the following work, we use logarithmic negativity, a widely used measure, to quantify quantum entanglement~\cite{vidal2002computable}. For an entanglement state described by a density matrix $\hat{\rho}$, the logarithmic negativity~\cite{vidal2002computable} is written as
\begin{equation}
\begin{aligned}
E_N(\hat{\rho})&= \log_2|| \hat{\rho}^{PT}||_1,
\label{EN}
\end{aligned}
\end{equation}
where
$PT$ is the partial transpose operation of $\hat{\rho}$ and $||X||_1= \mathrm{tr}|X| = \mathrm{tr}\sqrt{X^\dagger X}$ denotes the trace norm for an arbitrary operator $X$ with suitable dimensions.
For the entangled state $|q(\mathrm{R},\mathrm{B})\rangle$, its logarithmic negativity is calculated as $E_N(|q(\mathrm{R},\mathrm{B})\rangle \langle q(\mathrm{R},\mathrm{B}) |)$.

On the other hand, the performance of CV quantum teleportation is commonly evaluated by fidelity~\cite{furusawa2007quantum} indicating the distance between the input and the reconstructed quantum states written as
\begin{equation}
F_{CV}=\langle \psi_{\mathrm{E}}|\hat{\rho}_{out}|\psi_{\mathrm{E}}\rangle,
\label{fidelity}
\end{equation}
where $\hat{\rho}_{out}=|\psi_{out}(x_{-},p_{+})\rangle \langle \psi_{out}(x_{-},p_{+})|$ is the density matrix of the state reconstructed by Bob. The fidelity lies in between 0 and 1; i.e., $0 \leq F_{CV} \leq 1$. The closer the output state is to the input state, the closer $F_{CV}$ is to 1. In particular, when the input state is completely constructed by the Bob; i.e., $|\psi_{out}(x_{-},p_{+})\rangle=|\psi_{\mathrm{E}}\rangle$, $F_{CV}=1$. Furthermore, when using a random coherent state as the input state, the upper boundary in quantum teleportation is found to be $F_{classical}=0.5$~\cite{Braunstein2001quantum}. When $F_{CV}>F_{classical}$, the advantage of quantum entanglement is implied in achieving higher fidelity for quantum teleportation.

\section{CV quantum teleportation with a non-Markovian quantum channel}
In CV quantum teleportation, quantum information teleported from Alice to Bob via two channels: a classical communication channel transferring the Bell measurement results and a quantum communication channel sharing the entangled state. The latter would be sensitive to the influence of external environments, leading to the degrading performance of quantum teleportation. To attenuate undesirable perturbations from the environment and enhance the efficiency of quantum teleportation, we construct a non-Markovian quantum channel modeled by an augmented system. The augmented system and the quantum teleportation with a non-Markovian quantum channel will be introduced in this section.

\subsection{An augmented system}
\begin{figure}[ht]
\includegraphics[scale= 0.7]{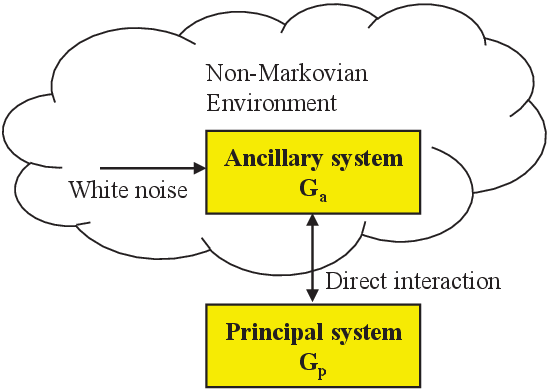}
\caption{\label{augmented system} Schematic plot of a general augmented system model for non-Markovian quantum systems.}
\end{figure}

In an augmented system model for non-Markovian quantum systems, ancillary systems are introduced to represent the internal modes of the non-Markovian environment, whose fictitious output carries a given quantum colored noise~\cite{xue2019modeling}. The augmented system is schematically plotted in Fig.~\ref{augmented system}. Besides the ancillary system driven by a white noise, the principal system is commonly the system of interest. Through direct interactions, they exchange energy and influence each other, resulting in the non-Markovian dynamics of the principal system.

We describe the augmented system using the so-called SLH description, where each subsystem can be described by a triple $G(S,L,\mathrm{H})$~\cite{gough2009series}. $S$ is a self-adjoint scattering matrix characterizing the scattering process of fields, $L$ is a coupling operator vector of the system with respect to the corresponding fields, and ${\mathrm H}$ is a Hamiltonian describing the internal energy of the system. Note that in the paper, we let $S=\mathrm{I}$ since the augmented model for the non-Markovian quantum channel in CV quantum teleportation does not involve scattering processes.

For the principal quantum system, since it does not couple with other fields as shown in Fig.~\ref{augmented system}, the coupling operator $L_p$ should be a null matrix; i.e., $L_p=O$. Therefore, the principal system should be described as,
\begin{equation}
G_p=({\rm I},O,{\rm H}_p),
\label{G_p_modify}
\end{equation}
where ${\rm H}_p$ is the Hamiltonian of the principal system defined on the Hilbert space $\mathcal{H}_p$.
Note that we use $O$ to denote a null matrix in this paper.

Similarly, the ancillary system is denoted as
\begin{equation}
G_a=({\mathrm I},L_a,{\mathrm H}_a),
\label{G_a}
\end{equation}
where ${\mathrm H}_a$ is the Hamiltonian of the ancillary system and $L_a$ is a coupling operator vector made up of a collection of coupling operators to quantum white noises. Both ${\mathrm H}_a$ and $L_a$ are defined on the Hilbert space $\mathcal{H}_a$, and the corresponding white noise is defined on the Fock space $\mathcal{F}_a$.

As for the energy exchange between the principal system and the ancillary system, it is represented by a direct interaction Hamiltonian,
\begin{equation}
{\rm H}_{pa}=i(c^\dagger z-z^\dagger c),
\label{H_pa}
\end{equation}
where the coupling operator $z$, defined on the Hilbert space $\mathcal{H}_p$, determines the influence of the principal system on the evolution of the ancillary quantum system, and the operator vector $c$, defined on the Hilbert space $\mathcal{H}_a$ of the ancillary system, represents the fictitious output operator, resulting in non-Markovian behavior of the principal system~\cite{xue2019modeling}.

According to the general quantum feedback network theory~\cite{gough2009series}, the augmented system composed of the principal system $G_p$, the ancillary system model $G_a$, and their interactions ~(\ref{H_pa}), is described in the SLH description as
\begin{eqnarray}
G_T=\left({\mathrm I}, \left( \begin{array}{c}
L_a\\
O
\end{array} \right), {\mathrm H}_p+{\mathrm H}_a+{\mathrm H}_{pa}\right).
\label{G_t}
\end{eqnarray}
Hence, the augmented system is defined on the augmented Hilbert space $\mathcal{H}_p \otimes \mathcal{H}_a \otimes \mathcal{F}_a$. The corresponding master equation is thus written as
\begin{eqnarray}
\dot{\hat{\rho}}(t) =&& -i[{\rm H}_p+{\rm H}_a+{\rm H}_{pa},\hat{\rho}(t)]+\mathcal{L}^\ast_{L_a}(\hat{\rho}(t)) 
\label{ms}
\end{eqnarray}
where $\hat{\rho}(t)$ is the the density matrix of the augmented system $G_T$, $\mathcal{L}^\ast_L$ is the adjoint of the Lindblad superoperator and calculated as $\mathcal{L}^\ast_L(\hat{\rho}(t)) = \frac{1}{2}[L\hat{\rho}(t),L^\dagger]+\frac{1}{2}[L,\hat{\rho}(t)L^\dagger]$.
The density matrix $\hat{\rho}_p(t)$ of the principal system can be obtained by performing the partial trace over the ancillary system,
\begin{equation}
\hat{\rho}_p(t)=\mathrm{tr}_a[\hat{\rho}(t)],
\label{partial}
\end{equation}
where ${\rm tr}_a[ \cdot ]$ is the partial trace operation on the ancillary system. Note that although the dynamics of the augmented system is Markovian as shown in Eq.~(\ref{ms}), the dynamics of the principal system is non-Markovian since it mutually exchanges energy with the ancillary system.

\subsection{CV Quantum teleportation with a non-Markovian quantum channel}
\begin{figure}[ht]
\includegraphics[scale= 0.55]{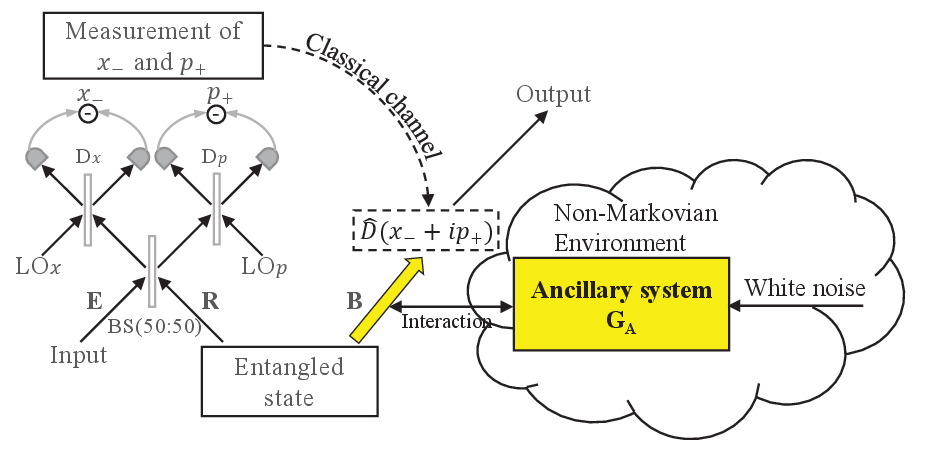}
\caption{\label{CVQT with non-Markovian} Schematic representation of CV quantum teleportation with a non-Markovian quantum channel. The displacement operation is represented by $\hat{D}(x_{-}+ip_{+})$. The yellow arrow represents the quantum channel in the CV quantum teleportation which is the principal system of the augmented system model.
}
\end{figure}

To investigate the performance of CV quantum teleportation with a non-Markovian quantum channel, we consider the channel is modelled by an augmented system as shown in Fig.~\ref{CVQT with non-Markovian}.
Compared with the traditional CV quantum teleportation in Section \uppercase\expandafter{\romannumeral2}, only the quantum communication channel is changed into a non-Markovian quantum channel and other parts are kept. In the augmented system model for the non-Markovian channel, the principal system is the original  quantum channel which is directly coupled to the ancillary system.

For the non-Markovian quantum channel, we follow the augmented system approach and directly obtain its master equation from Eq.~(\ref{ms}) as
 \begin{equation}
 \dot{\hat{\rho}}(t) = -i[{\rm H}_\mathrm{B}+{\rm H}_\mathrm{A}+{\rm H}_\mathrm{BA},\hat{\rho}(t)]+\mathcal{L}^\ast_{L_\mathrm{A}}(\hat{\rho}(t)),
 \label{msB}
\end{equation}
where ${\rm H}_\mathrm{B}$ and ${\rm H}_\mathrm{A}$ are the Hamiltonian of the principal system and the ancillary system, respectively.
Because a quantum state does not evolve in an ideal quantum channel, we ignore itself dynamics such that ${\rm H}_\mathrm{B}=\omega_b\mathrm{I}$, where $\omega_b$ is the angular frequency.
The interaction Hamiltonian ${\rm H}_\mathrm{BA}$ describes the energy exchange between the states in the quantum channel and the environment.

In the non-Markovian quantum channel, mode $\mathrm{B}$ of the entangled state $|q(\mathrm{R},\mathrm{B})\rangle$
evolves during the transmission from Alice to Bob. Although mode $\mathrm{B}$ obeys a non-Markovian dynamic in quantum teleportation, mode $\mathrm{R}$ would be affected due to the entanglement between them.
Therefore, the master equation for the entangled state $|q(\mathrm{R},\mathrm{B})\rangle$ affected by the non-Markovian quantum channel can be rewritten as,
\begin{eqnarray}
\dot{\hat{\rho}}(t) =&& -i[{\rm H}_\mathrm{P}+{\rm H}_\mathrm{A}+{\rm H}_\mathrm{BA},\hat{\rho}(t)]+\mathcal{L}^\ast_{L_\mathrm{A}}(\hat{\rho}(t)) 
\label{ms_RB}
\end{eqnarray}
where ${\rm H}_\mathrm{P}$ is the Hamiltonian for the entangled state $|q(\mathrm{R},\mathrm{B})\rangle$.
It can be written as ${\mathrm H}_\mathrm{P} = {\mathrm H}_\mathrm{R} \otimes \mathrm{I}_{N_\mathrm{B}}+ \mathrm{I}_{N_\mathrm{R}} \otimes {\rm H}_\mathrm{B}$, where the subscript $N_\mathrm{R}$ and $N_\mathrm{B}$ denote the dimensions of mode $\mathrm{R}$ and mode $\mathrm{B}$, respectively. Since mode $\mathrm{R}$ does not involve an evolution induced by itself in the process of teleportation, the Hamiltonian operator can be written as ${\mathrm H}_\mathrm{R}=\omega_r \mathrm{I}$, where $\omega_r$ is the angular frequency.

Define the transit time as $t=t_e-t_0$, where $t_0$ is the time instant when mode $\mathrm{B}$ starts to be transmitted from Alice to Bob in the quantum non-Markovian channel, and $t_e$ is the time instant when the state is received by Bob.
For simplicity, we set $t_0=0$, then $t=t_e$.
The initial density matrices of the principal system and the ancillary system are denoted as $\hat{\rho}_\mathrm{RB}(0)=\hat{\rho}_{q(\mathrm{R},\mathrm{B})}= |q(\mathrm{R},\mathrm{B})\rangle \langle q(\mathrm{R},\mathrm{B})|$ and $\hat{\rho}_\mathrm{A}(0)$, respectively, and then
the initial density operator of the augmented system in the master Eq.~(\ref{ms_RB}) is $\hat{\rho}(t_0)=\hat{\rho}(0)=\hat{\rho}_\mathrm{RB}(0) \otimes \hat{\rho}_\mathrm{A}(0)$. With the method described in Appendix A, the density matrix $\hat{\rho}(t)$ of the augmented system at any transit time $t$ can be calculated. Then, the density matrix $\hat{\rho}_\mathrm{RB}(t)$ of the entangled state
at any transit time $t$ can be obtained by tracing over the ancillary system as
\begin{equation}
\hat{\rho}_\mathrm{RB}(t)=\hat{\rho}_\mathrm{P}(t)=\mathrm{tr}_\mathrm{A}[\hat{\rho}(t)].
\label{principal_solution}
\end{equation}
Note that we have established an augmented system framework for investigation of non-Markovian effects on CV quantum teleportation.
With the above model, substituting Eq.~(\ref{principal_solution}) into Eq.~(\ref{EN}), we can calculate the corresponding logarithmic negativity of the entangled state at the transit time $t$ as
\begin{equation}
E_N(\hat{\rho}_{q(\mathrm{R},\mathrm{B})},t)=E_N(\hat{\rho}_\mathrm{RB}(t))
=\log_2|| \hat{\rho}_\mathrm{RB}(t)^{PT}||_1.
\end{equation}

After the propagation in the non-Markovian channel, the initial entangled state $\hat{\rho}_{q(\mathrm{R},\mathrm{B})}$ evolves into $\hat{\rho}_\mathrm{RB}(t)$. Modes $\mathrm{R}$ and $\mathrm{B}$ reach Alice and Bob, respectively. Then Alice mixes mode $\mathrm{R}$ with the input state denoted by $\hat{\rho}_\mathrm{E}$ through a half beam splitter and performs two sets of balanced homodyne detections on the ensuing state.
We assume both the beam splitter and the balanced homodyne detector are lossless. The measurement is a projection on the maximally entangled basis~\cite{kitagawa2006entanglement}
\begin{equation}
|\Pi(x_{-},p_{+})\rangle_\mathrm{ER}=\frac{1}{\sqrt{\pi}}\int _{-\infty}^{\infty} dye^{2ip_{+}y}|x_{-}+y\rangle_\mathrm{E}|y\rangle_\mathrm{R},
\end{equation}
which is equivalent to Eq.~(\ref{measure_basis}), written in the Fock basis.
After the measurement, Alice gets the result $(x_{-},p_{+})$ and mode $\mathrm{B}$ at Bob is given as
\begin{equation}
\begin{split}
\hat{\sigma}_\mathrm{B}(x_{-},p_{+},\hat{\rho}_{q(\mathrm{R},\mathrm{B})},t)= \qquad \qquad  \qquad \qquad \quad \quad \quad \\
\frac{{_\mathrm{ER}} \langle\Pi(x_{-},p_{+})|[\hat{\rho}_\mathrm{E}\otimes \hat{\rho}_\mathrm{RB}(t)]|\Pi(x_{-},p_{+})\rangle_\mathrm{ER}}{P(x_{-},p_{+},\hat{\rho}_{q(\mathrm{R},\mathrm{B})},t)},
\end{split}
\end{equation}
where
\begin{eqnarray}
P(x_{-},p_{+},\hat{\rho}_{q(\mathrm{R},\mathrm{B})},t)=&&\mathrm{tr}_\mathrm{B}\{{_\mathrm{ER}} \langle\Pi(x_{-},p_{+})|[\hat{\rho}_\mathrm{E}\otimes \hat{\rho}_\mathrm{RB}(t)] \nonumber \\
&& |\Pi(x_{-},p_{+})\rangle_\mathrm{ER}\}
\end{eqnarray}
is the probability at transit time $t$~\cite{kitagawa2006entanglement}. Finally, Bob reconstructs the input state $\hat{\rho}_\mathrm{E}$ using the information $(x_{-},p_{+})$, transferred through the classical channel. The density matrix of the output state becomes
\begin{equation}
\begin{split}
\hat{\rho}_{out}(x_{-},p_{+},\hat{\rho}_{q(\mathrm{R},\mathrm{B})},t)= \qquad \quad  \quad \qquad \qquad \qquad \\
\hat{D}_\mathrm{B}(x_{-}+ip_{+}) \hat{\sigma}_\mathrm{B}(x_{-},p_{+},\hat{\rho}_{q(\mathrm{R},\mathrm{B})},t) \hat{D}^\dagger_\mathrm{B}(x_{-}+ip_{+}),
\end{split}
\end{equation}
where the displacement operator $\hat{D}_\mathrm{B}(x_{-}+ip_{+})=\mathrm{exp}[(x_{-}+ip_{+}) \hat{a}^\dagger-(x_{-}-ip_{+})\hat{a}]$ with the annihilation operator $\hat{a}$ and the creation operator $\hat{a}^\dagger$ for mode $\mathrm{B}$.

For a single quantum teleportation with a Bell measurement result $(x_{-},p_{+})$, the fidelity is given by
\begin{equation}
F(x_{-},p_{+},\hat{\rho}_{q(\mathrm{R},\mathrm{B})},t)=\mathrm{tr}[\hat{\rho}_\mathrm{E}\hat{\rho}_{out}(x_{-},p_{+},\hat{\rho}_{q(\mathrm{R},\mathrm{B})},t)].
\end{equation}
By averaging over all measurements, the average fidelity $\bar{F}(t)$ at a transit time $t$ can be calculated as
\begin{eqnarray}
\bar{F}(\hat{\rho}_{q(\mathrm{R},\mathrm{B})},t) =&& \int _{-\infty}^{\infty} dx_{-} \int _{-\infty}^{\infty} dp_{+} P(x_{-},p_{+},\hat{\rho}_{q(\mathrm{R},\mathrm{B})},t) \nonumber \\
&& F(x_{-},p_{+},\hat{\rho}_{q(\mathrm{R},\mathrm{B})},t).
\label{f_av}
\end{eqnarray}
\section{Effects of non-markovianity on CV quantum teleportation}
As we have introduced the non-Markovian quantum channel for CV quantum teleportation, it is important to investigate how the non-Markovianity of the channel affects the performance of the teleportation. In this section, the measure of the non-Markovianity is introduced at first. Then the effects of the non-Markovian dynamics of quantum channels are analyzed based on the CV quantum teleportation with a non-Markovian quantum channel disturbed by quantum Lorentzian noise. Moreover, relevant results are extended from Lorentzian noise to the environment with a rational power spectral density for generality.
\subsection{Measure of non-Markovianity}
In this paper, we simply adopt the Breuer-Laine-Pillo (BLP) quantifier to measure the non-Markovianity of the quantum channel, which depends on the behavior of the trace distance in Ref.~\cite{breuer2009measure}. Concretely, the trace distance for two quantum states $\hat{\rho}_1$ and $\hat{\rho}_2$ is defined as
\begin{equation}
\mathcal{D}(\hat{\rho}_1,\hat{\rho}_2) =\mathrm{tr}|\hat{\rho}_1-\hat{\rho}_2|/2.
\label{tr}
\end{equation}
where $|X|=\sqrt{X^\dagger X}$ for an arbitrary operator $X$ with suitable dimensions. It satisfies a contractive property for any completely positive (CPT) map $\Phi$, namely,
\begin{equation}
\mathcal{D}(\Phi(\hat{\rho}_1),\Phi(\hat{\rho}_2)) \leq \mathcal{D}(\hat{\rho}_1,\hat{\rho}_2).
\label{non2}
\end{equation}
The inequality~(\ref{non2}) holds in the Markovian process because of the divisibility property. The monotonic decrease implies the flow of information from the system of interest to the environment. Conversely, the key characteristic of non-Markovianity is the inverse flow of information from the environment to the system, leading to the failure of inequality~(\ref{non2}) for certain times. Therefore, in Ref.~\cite{breuer2009measure}, non-Markovianity is defined as the physical process where
\begin{equation}
\nu(t,\hat{\rho}_{1,2}(0)) = \frac{d}{dt} \mathcal{D}(\hat{\rho}_1(t),\hat{\rho}_2(t))>0
\end{equation}
exists for certain times.

To measure the total amount of information flowing from the environment to the system,
\begin{equation}
\mathcal{N}(\Phi) = \max_{\hat{\rho}_{1,2}(0)} \int_{\nu>0}dt \nu(t,\hat{\rho}_{1,2}(0))
\label{BLP}
\end{equation}
is used to quantify the non-Markovianity of a quantum process for the quantum process $\Phi$. Here, $\mathcal{N}(\Phi)$ takes the time-integration on all time intervals in which $\nu > 0$, namely the accumulation of the increased amount of the trace distance of two quantum states in the quantum evolution, and then chooses the maximum value over all possible pairs of initial states.
In the following subsection, we will use the BLP quantifier $\mathcal{N}$ as the measure of the non-Markoviantiy of the quantum channel in CV quantum teleportation.

\subsection{Lorentzian noise environment}
Quantum Lorentzian noise is a common quantum colored noise, which has been observed in two-level systems~\cite{tubsrinuan2021probing,burnett2019decoherence} and detector-preamplifier setups~\cite{pullia2004time}. Moreover, Lorentzian noise power spectral density has also been taken into consideration in the investigation of quantum communication channels, such as optical fibers~\cite{drummond2001quantum} and quantum channel models~\cite{etxezarreta2021time}.
In this subsection, we focus on the CV quantum teleportation with a non-Markovian quantum channel disturbed by quantum Lorentzian noise. The corresponding quantum channel is described at first, and then the effects of non-Markovianity on both the logarithmic negativity of the entangled state and the average fidelity of quantum teleportation are investigated.
\subsubsection{Quantum channels disturbed by quantum Lorentzian noise}
For a quantum Lorentzian noise with a spectrum width $\frac{\gamma_0}{2}$ and a characteristic frequency $\omega_0$,
its power spectral density is written as
\begin{equation}
S_0(\omega)=\frac{\frac{\gamma_0^2}{4}}{\frac{\gamma_0^2}{4}+(\omega-\omega_0)^2}.
\label{L_noise}
\end{equation}
It can be generated by a one-mode quantum harmonic oscillator driven by quantum white noise. Concretely, its Hamiltonian ${\rm H}_\mathrm{A}$ and coupling operator $L_\mathrm{A}$ are written as
\begin{equation}
{\rm H}_\mathrm{A}= \omega_0\hat{a}_0^\dagger \hat{a}_0,\quad L_\mathrm{A}=\sqrt{\gamma_0} \hat{a}_0,
\end{equation}
where $\hat{a}_0$ and $\hat{a}_0^\dagger$ are the annihilation and creation operators of the ancillary system, respectively~\cite{xue2019modeling}. Parameters $\omega_0$ and $\gamma_0$ are the angular frequency and damping rate of the ancillary system, respectively.

Since in principle the dimension of the oscillator is infinite-dimensional, we have to truncate it to a finite dimension $N_\mathrm{A}$ for approximation. Thus the state of the ancillary is a ground state system $|0\rangle_{N_\mathrm{A}}$. In the direct interaction $\mathrm{H}_\mathrm{BA}$, the fictitious output is given as $c=-\frac{\sqrt{\gamma_0}}{2} \hat{a}_0$, which takes Lorentzian spectrum and the coupling operator is $z=\sqrt{\kappa_0} \hat{a}_\mathrm{B}$, where $\hat{a}_\mathrm{B}$ is the the annihilation operator of the principal system and $\kappa_0$ is the coulping strength. To ensure that the ancillary system can effectively affect the channel state, we only consider the resonant case with zero detuning; i.e., $\omega_0=\omega_b$.
Note that when $\gamma_0 \rightarrow \infty$, quantum Lorentzian noise modeled by the ancillary system reduces to quantum white noise and the non-Markovian quantum channel reduces to a Markovian quantum channel. Meanwhile, the evolution of the principal system is described by the following master equation~\cite{xue2015quantum},
\begin{equation}
\dot{\hat{\rho}}_\mathrm{B}(t)=-i[\mathrm{H}_\mathrm{B},\hat{\rho}_\mathrm{B}(t)]+\mathcal{L}^\ast_{z}(\hat{\rho}_\mathrm{B}(t)),
\label{M_ms_one}
\end{equation}
where $\mathcal{L}^\ast_{z}(\hat{\rho}_\mathrm{B}(t))= \frac{1}{2}[z\hat{\rho}_\mathrm{B}(t),z^\dagger]+\frac{1}{2}[z,\hat{\rho}_\mathrm{B}(t)z^\dagger]$. Obviously, the dynamics of the corresponding Markovian quantum channel is mainly determined by the parameter $\kappa_0$.

\begin{figure}[ht]
\includegraphics[scale=0.28]{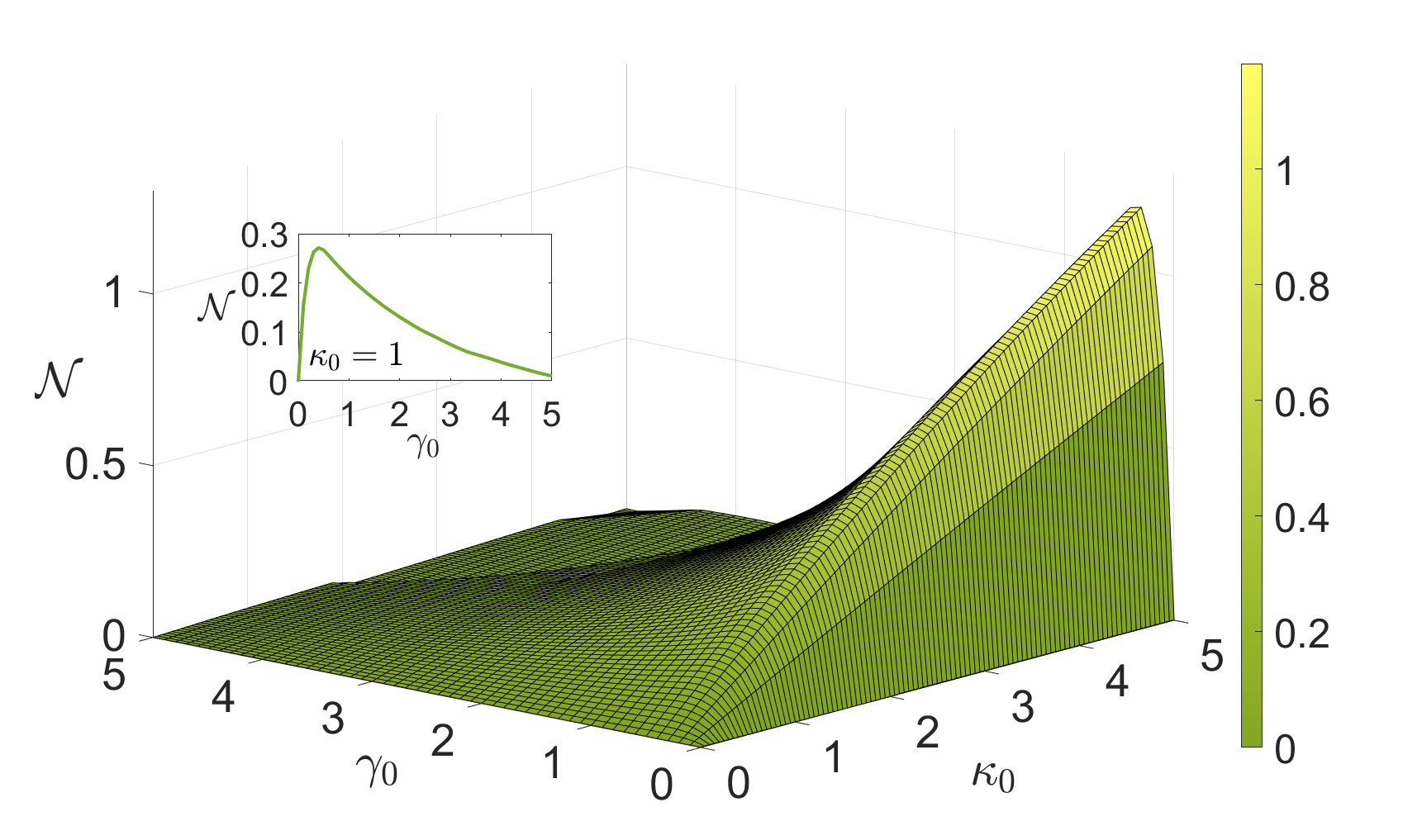}
\caption{\label{parameter_rk} The relationship between the non-Markovianity of a quantum channel and the parameters $\gamma_0$ and $\kappa_0$ of the ancillary system.
The $x$ axis is the the damping rate $\gamma_0$, the $y$ axis is the coupling strength $\kappa_0$, and the $z$ axis is the BLP quantifier $\mathcal{N}$. The insert subplot represents the non-Markovianity $\mathcal{N}$ as a function of $\gamma_0$ when $\kappa_0=1$. In the subplot, the non-Markovianity of the quantum channel increases and then decreases as the parameter $\gamma_0$ increases. }
\end{figure}

The non-Markovian dynamics of the quantum channel modeled by an augmented system results from the direct interaction between the ideal quantum channel and the ancillary system. In the above model, the non-Markovianity of the quantum channel is affected by two parameters, the damping rate $\gamma_0$ and the direct coupling strength $\kappa_0$. Fig.~\ref{parameter_rk} shows how the BLP quantifier $\mathcal{N}$ of a quantum channel varies with the two parameters. Note that $\gamma_0$ determines the bandwidth of the Lorentzian spectrum, as shown in Eq.~(\ref{L_noise}). And $\kappa_0$ is closely related to amplitude of the Lorentzian spectrum modeled by the ancillary system, as explained in Ref.~\cite{xue2019modeling}.

With the increase of the parameter $\gamma_0$, the non-Markovianity of the quantum channel shows a tendency to increase and then decrease, as shown in the insert subplot in Fig.~\ref{parameter_rk}. Within a certain range, e.g., $0 < \gamma_0<0.3 $, as $\gamma_0$ increases, a broader bandwidth of the noise spectrum can enhance the non-Markovian effect. While, as $\gamma_0$ keeps increasing, e.g., $0.4 < \gamma_0<5$, the bandwidth is getting broader and broader, and the spectrum gradually tends to a flatten one. At the same time, the impact brought about also gradually decreases, and subsequently, the non-Markovian dynamics gradually weaken or even disappear. In addition, as the coupling strength $\kappa_0$ increases, the amplitude of the Lorentzian noise spectrum increases, and thus intensifying the non-Makrovianity of the quantum channel as predicted. Moreover, the amplitude of the non-Markovianity of the quantum channel disturbed by the Lorentzian noise is determined by both $\gamma_0$ and $\kappa_0$, which is indicated by the slight fluctuations that exist in the downward tendency of the non-Markovianity with the increase of the damping rate $\gamma_0$.
\subsubsection{The effects of non-Markovianity on the entangled state}
As introduced, it is one mode of the entangled state that propagates in the non-Markovian quantum channel, so the entanglement is directly affected by the non-Markovian dynamics. Since the entanglement plays a critical role in the quantum teleportation and determines the performance of the system, it is necessary to investigate the effect of the non-Markovian channel on the entangled state to better understand the influence on quantum teleportation.

\begin{figure}[ht]
\includegraphics[scale=0.26]{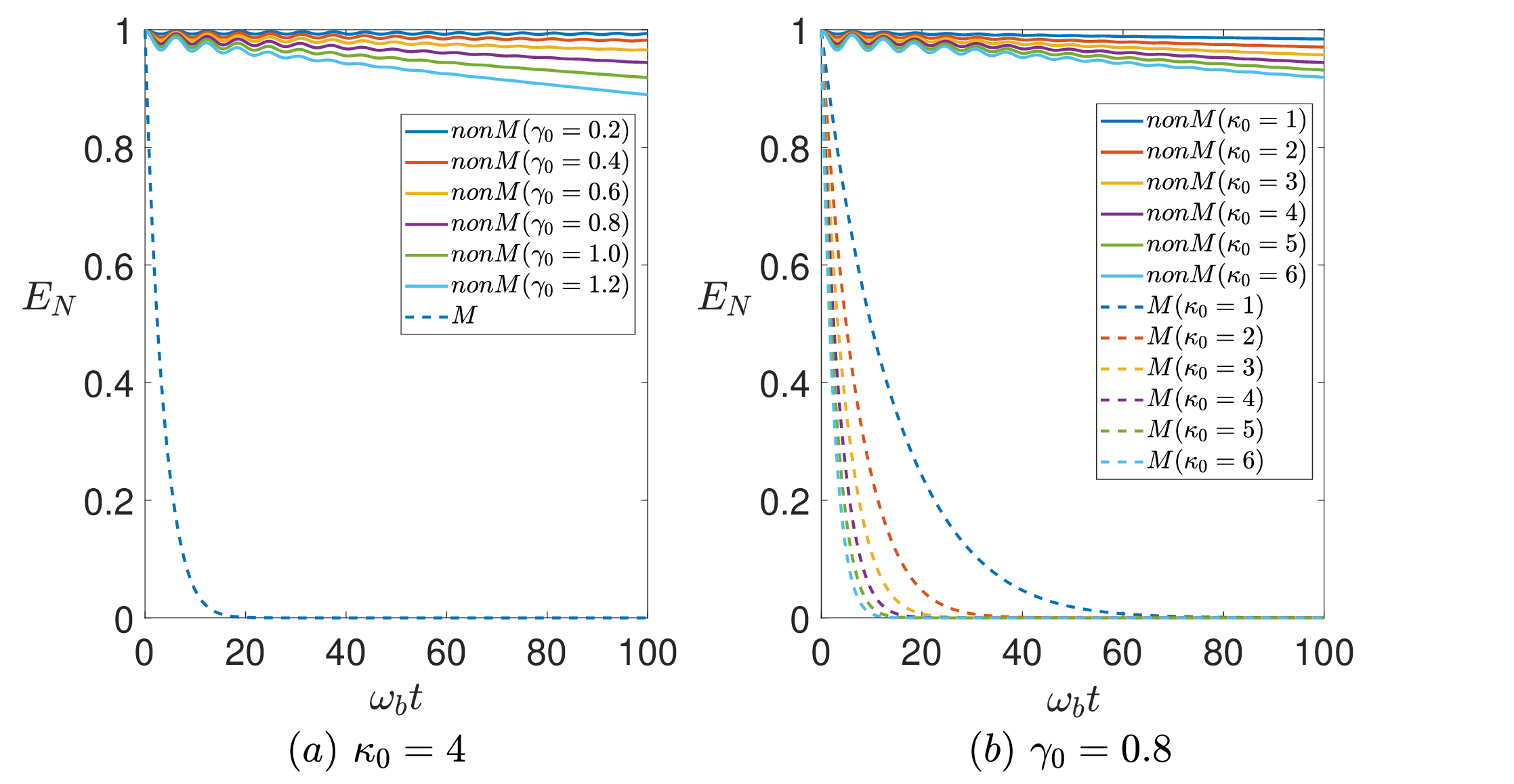}
\caption{\label{parameter_r_k_rs}($a$) Variation of the logarithmic negativity $E_N$ with different $\gamma_0$ where $\kappa_0=4$. (b) Variation of the logarithmic negativity $E_N$ with different $\kappa_0$ where $\gamma_0=0.8$.}
\end{figure}

To analyze the teleportation process, we choose a squeezing parameter $r=0.346$ $(-3$ $\mathrm{dB})$ for the two-mode squeezed vacuum state.
The variation of the logarithmic negativity of the two-mode squeezed vacuum state transferred in the non-Markovian quantum channel disturbed by Lorentzian noise with different parameters are plotted in Fig.~\ref{parameter_r_k_rs}, where the cases of corresponding Markovian quantum channels are also considered. Here, the coupling strength $\kappa_0$ is set to be 4 in Fig.~\ref{parameter_r_k_rs}($a$) and the damping rate $\gamma_0$ is given as 0.8 in Fig.~\ref{parameter_r_k_rs}($b$).
Note that since the corresponding Markovian quantum channel is determined by $\kappa_0$, in Fig.~\ref{parameter_r_k_rs}($a$), the corresponding Markovian quantum channels are the same for the non-Markovian quantum channels with the same $\kappa_0$.

Under the disturbance of the environment, the entanglement of the two-mode squeezed vacuum state decays in both non-Markovian and Markovian quantum channels as shown in Fig.~\ref{parameter_r_k_rs}. In the Markovian quantum channels, the logarithmic negativity of the entangled state rapidly decays. In contrast, that is well maintained in the non-Markovian quantum channels and even shows a slight fluctuation during the decay. In addition, as $\gamma_0$ or $\kappa_0$ increases, $E_N$ of the entangled state decays faster in transmission but still much more slowly than that for Markovian quantum channels.

\begin{figure}[ht]
\includegraphics[scale=0.225]{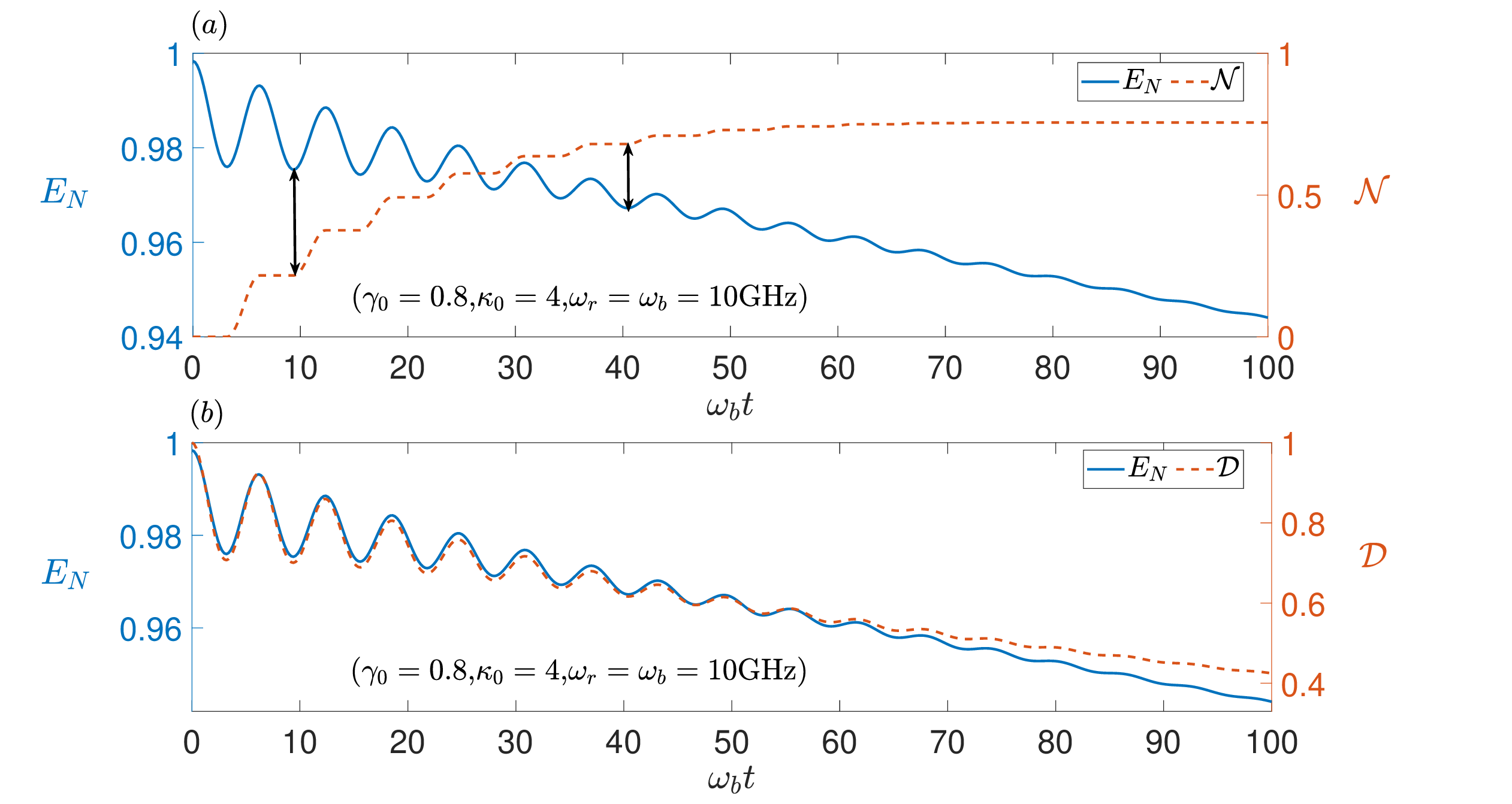}
\caption{\label{ln_nonM}(Color online) ($a$) The logarithmic negativity $E_N$ (blue full curve) and the BLP quantifier $\mathcal{N}$ (red dashed curve) as functions of the dimensionless time $\omega_bt$ in the non-Markovian quantum channel. ($b$) The logarithmic negativity $E_N$ (blue full curve) and the trace distance $\mathcal{D}$ (red dashed curve) as functions of the dimensionless time $\omega_bt$ in the non-Markovian quantum channel.
}
\end{figure}

The relationship between the non-Markovianity of the quantum channel and the logarithmic negativity of the entangled state is further plotted in Fig.~\ref{ln_nonM}, which is based on the non-Markovian quantum channel disturbed by the Lorentzian noise with the parameters $\gamma_0=0.8$ and $\kappa_0=4$.
Fig.~\ref{ln_nonM}($a$) shows the evolution of logarithmic negativity $E_N$ of the entangled state and the BLP quantifier $\mathcal{N}$ in the non-Markovian quantum channel. Similarly, the relationship between the evolution of logarithmic negativity $E_N$  and the trace distance $\mathcal{D}$ of two quantum states is shown in Fig.~\ref{ln_nonM}($b$). Each fluctuating rebound of the trace distance $\mathcal{D}$ corresponds to an increase of the BLP quantifier $\mathcal{N}$. From the two subplots, it is clear that the increase of the non-Markovianity of the quantum channel can efficiently suppress the decline of the entanglement of the resource state and even brings about recovery. In addition, in Fig.~\ref{ln_nonM}($b$), the two curves show a near coincidence in frequency and amplitude of fluctuation. The consistency of the fluctuating frequency strongly illustrates the dependence of the entangled state on the non-Markovian dynamics of the quantum channel, demonstrating that the attenuation of the logarithmic negativity of the entangled state stems from the effect of non-Markovian dynamics of the quantum channel. The consistency of the fluctuating amplitude indicates the close correlation between the efficiency of entanglement revival and the amount of information backflow.

\begin{figure*}
\centering
\includegraphics[scale=0.33]{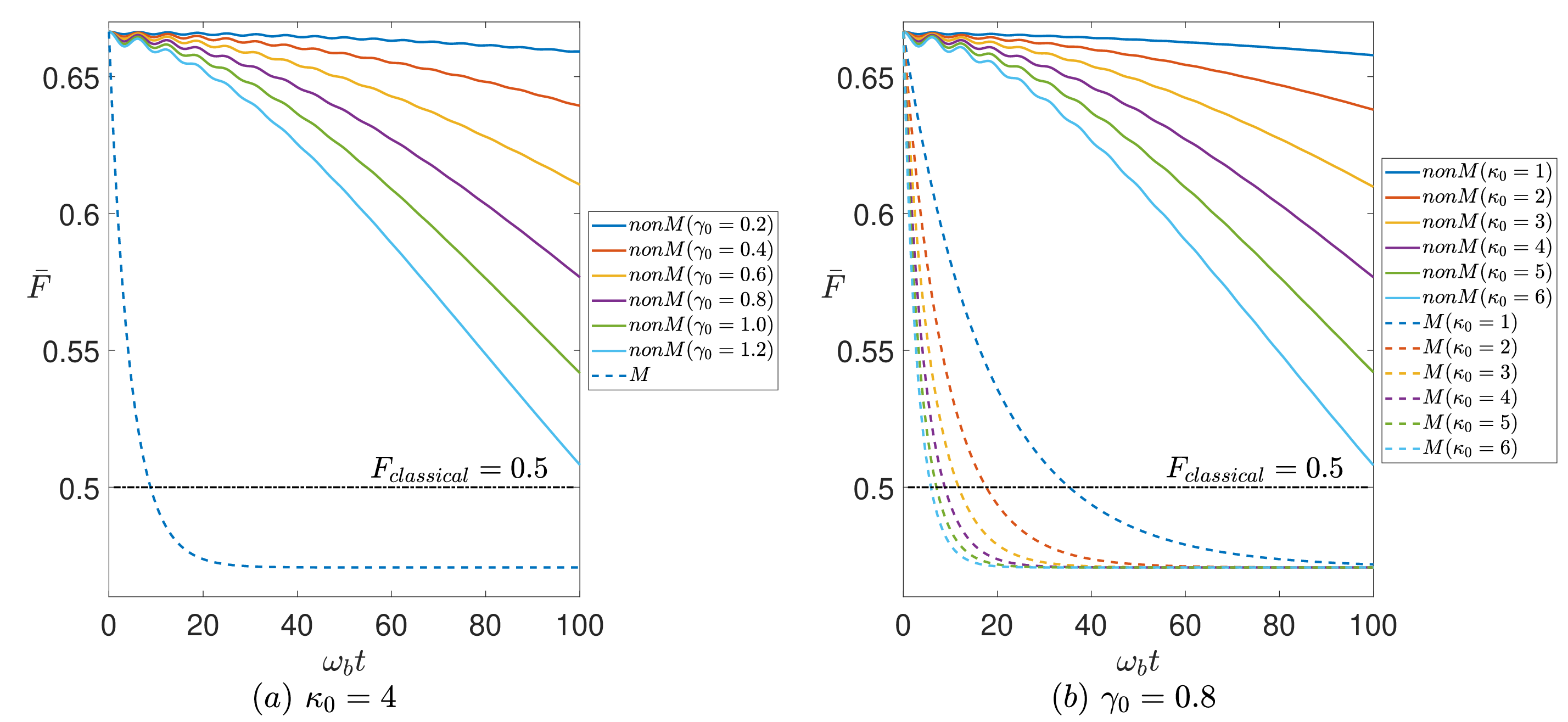}
\caption{\label{fidelity_parameter_r_k}($a$) Variation of the average fidelity $\bar{F}$ with different $\gamma_0$ where $\kappa_0=4$. ($b$) Variation of the average fidelity $\bar{F}$ with different $\kappa_0$ where $\gamma_0=0.8$. The horizontal dash-dotted line represents the upper boundary of classical teleportation, $F_{classical}=0.5$.}
\end{figure*}
\subsubsection{The effects of non-Markovianity on CV quantum teleportation}
The indispensable role of the entangled state in CV quantum teleportation indicates that the effects of the non-Markovian channel on the entangled state inevitably influence the performance of CV quantum teleportation. In this subsection, we will analyze the effects of the non-Markovian dynamics of the quantum channel on the average fidelity of CV quantum teleportation.

We consider the case that the input state for quantum teleportation is a coherent state. The coherent state is commonly represented by $|\alpha\rangle=\hat{D}(\alpha)|0\rangle$, where $|0\rangle$ is a vacuum state and $\hat{D}(\alpha)$ is a displacement operator written as $\hat{D}(\alpha)=\mathrm{exp}(\alpha \hat{a}^\dagger-\alpha^\ast\hat{a})$, where $\hat{a}$ and $\hat{a}^\dagger$ are the annihilation and creation operators, respectively. The input coherent state with a complex parameter $\alpha_\mathrm{E}$ is written as $|\psi\rangle_\mathrm{E}=\hat{D}(\alpha_\mathrm{E})|0\rangle$ and thus corresponding density matrix is written as $\hat{\rho}_\mathrm{E}=|\psi\rangle_\mathrm{E}\langle \psi|$.

The influence of the non-Markovianity of the quantum channel on the quantum teleportation is shown in Fig.~\ref{fidelity_parameter_r_k}.
Fig.~\ref{fidelity_parameter_r_k} reveals that the decay of the average fidelity of the quantum teleportation is effectively suppressed by the non-Markovian dynamics of the quantum channel, compared with the corresponding Markovian quantum channel. Especially, in the early stage of information transmission, e.g., $0 \leq \omega_bt \leq 20$, Fig.~\ref{fidelity_parameter_r_k}($a$) and ($b$) clearly show that under the influence of the non-Markovian quantum channel, $\bar{F}$ presents fluctuating recovery and holds relatively stable during decay. While in the corresponding Markovian quantum channel, $\bar{F}$ decays at a high rate, which is consistent with the rapid decrease in the logarithmic negativity of the entangled state shown in Fig.~\ref{parameter_r_k_rs}. As the transmission time $\omega_bt$ increases, e.g., $40 \leq \omega_bt \leq 100$, $\bar{F}$ of the quantum teleportation with a non-Markovian quantum channel shown in Fig.~\ref{fidelity_parameter_r_k}($a$) and ($b$) decreases, but is still better than that of the quantum teleportation with the corresponding Markovian quantum channel.
Additionally, Fig.~\ref{fidelity_parameter_r_k}($a$) and ($b$) show that in the case of non-Markovian quantum channels, $\bar{F}> F_{classical}$ indicates that the entanglement remains the critical ingredient.
Furthermore, the decay rate of $\bar{F}$ gradually accelerated with the increase of $\gamma_0$ or $\kappa_0$, as shown in Fig.~\ref{fidelity_parameter_r_k}($a$) and ($b$), respectively, which is similar to the variation of $E_N$ of the entangled state as shown in Fig.~\ref{parameter_r_k_rs}($a$) and ($b$).

\begin{figure}[ht]
\includegraphics[scale=0.24]{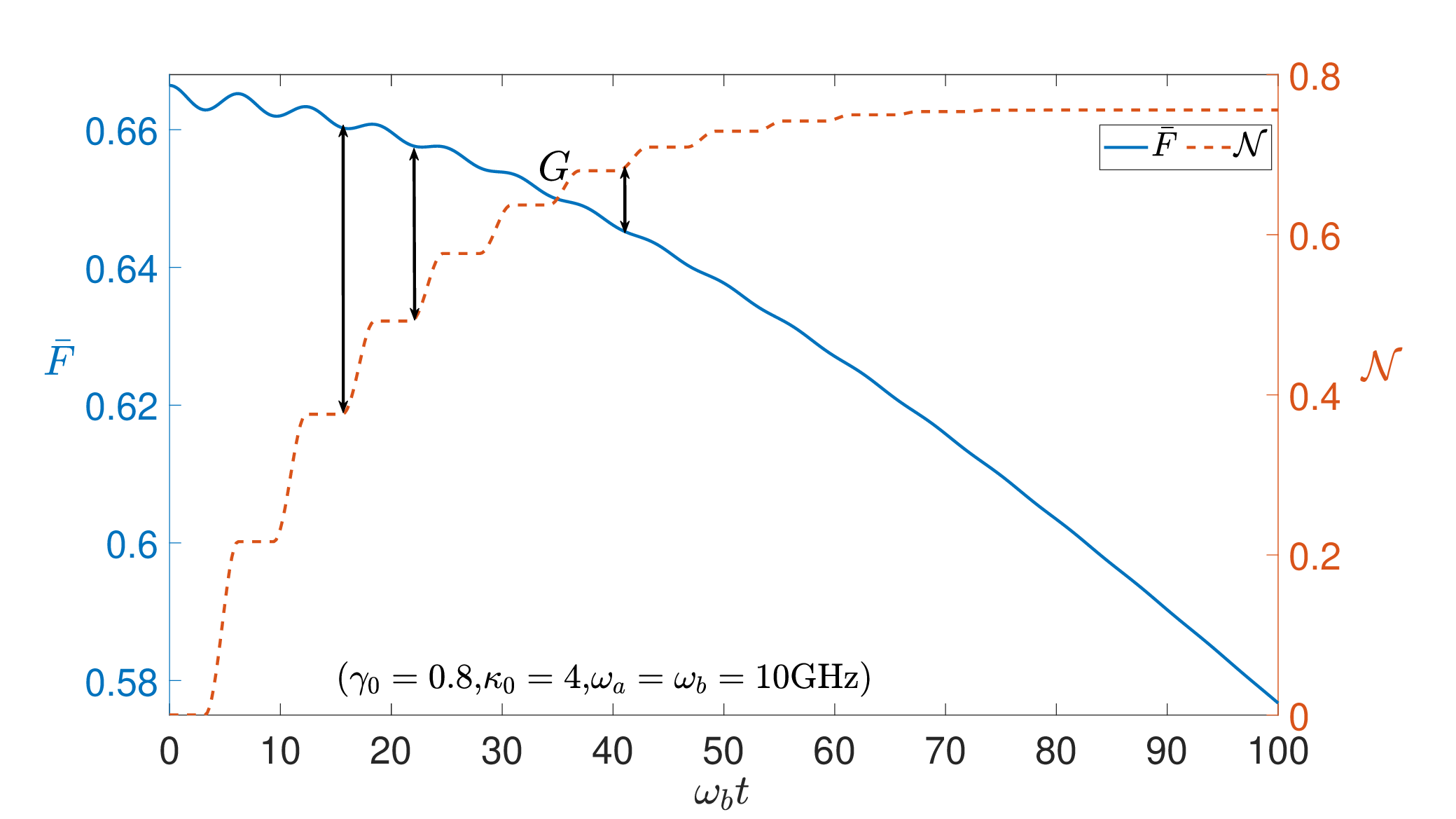}
\caption{\label{nonM_F_dsum}(Color online) The average fidelity $\bar{F}$ (blue full curve) and the BLP quantifier $\mathcal{N}$ (red dashed curve) as functions of the dimensionless time $\omega_bt$ in the non-Markovian quantum channel. And the letter $G$ represents the intersection of two curves.
}
\end{figure}
The relationship between the average fidelity and the non-Markovianity of the quantum channel is plotted in Fig.~\ref{nonM_F_dsum}, which is based on the same non-Markovian quantum channel with parameters $\gamma_0=0.8$ and $\kappa_0=4$. Similar to the effect of the non-Markovianity of the quantum channel on the entangled state shown in Fig.~\ref{ln_nonM}, the increase in non-Markovianity $\mathcal{N}$ per time suppresses the decline of the average fidelity $\bar{F}$ and even brings about recovery. Besides, the magnitude of each increase in non-Markovianity is closely related to the intensity of the effect on quantum teleportation. Concretely, with the intersection $G$ as the demarcation point, the non-Markovianity on the left side exhibits a higher magnitude of each increase and corresponds to a greater recovery of the average fidelity. In contrast, the non-Markovianity on the right increases less and corresponds to less recovery, or even shows only suppression. It also explains well that through the non-Markovian quantum channel, the average fidelity $\bar{F}$
is relatively stable in the early transmission stage, e.g., $0 \leq \omega_bt \leq 20$, and then deteriorates as the transmission time increases.

\subsection{The environment with a rational power spectral density}
A variety of quantum colored noise exists in actual physical systems. The ancillary system can model quantum colored noise with a rational power spectral density~\cite{xue2019modeling}. In this section, we take another type of quantum colored noise, two-Lorentzian noise, as an example to illustrate the generality of the augmented system framework for CV quantum teleportation.

The power spectral density of two-Lorentzian noise is described as
\begin{equation}
S(\omega)=\frac{(\frac{\gamma_1}{2})^2}{(\frac{\gamma_1}{2})^2+(\omega-\omega_1)^2}+\frac{(\frac{\gamma_2}{2})^2}{(\frac{\gamma_2}{2})^2+(\omega-\omega_2)^2},
\end{equation}
where $\frac{\gamma_1}{2}$ and $\frac{\gamma_2}{2}$ are the widths of the spectrum,
and $\omega_1$ and $\omega_2$ are the center frequencies. It can be modeled by two quantum harmonic oscillators, and the master equation is written as
\begin{equation}
\dot{\hat{\rho}}(t)=-i[\mathrm{H}_\mathrm{B}+\sum_{k=1}^2 (\mathrm{H}_{\mathrm{A}_k}+\mathrm{H}_{\mathrm{B}{\mathrm{A}_k}}),\hat{\rho}(t)]+\sum_{k=1}^2 \mathcal{L}^\ast_{L_{\mathrm{A}_k}}(\hat{\rho}(t)).
\end{equation}
The Hamiltonian of the $k$-th ancillary system $\mathrm{H}_{\mathrm{A}_k}=\omega_k \hat{a}_k^\dagger \hat{a}_k$, where $\hat{a}_k$ and $\hat{a}_k^\dagger$ are the annihilation and creation operators of the $k$-th ancillary system, respectively. The interaction Hamiltonian between the principal system and the $k$-th ancillary system is $\mathrm{H}_{\mathrm{B}\mathrm{A}_k}=i(c_k^\dagger z_k-z_k^\dagger c_k)$ with the operator vector $c_k=-\frac{\sqrt{\gamma_k}}{2}\hat{a}_k$ and the coupling operator $z_k=\sqrt{\kappa_k}\hat{a}_k$. The coupling operator of the $k$-th ancillary system is $L_{\mathrm{A}_k}=\sqrt{\gamma_k}\hat{a}_k$. Note that when $\gamma_k \rightarrow \infty$, $k=1,2$, two-Lorentzian noise modeled by the ancillary system reduces to white noise and the non-Markovian quantum channel reduces to a Markovian quantum channel. The evolution of the principal system is given by the following master equation~\cite{xue2015quantum},
\begin{equation}
\dot{\hat{\rho}}_\mathrm{B}(t)=-i[\mathrm{H}_\mathrm{B},\hat{\rho}_\mathrm{B}(t)]+\mathcal{L}^\ast_{z_1}(\hat{\rho}_\mathrm{B}(t))+\mathcal{L}^\ast_{z_2}(\hat{\rho}_\mathrm{B}(t)),
\label{M_ms_two}
\end{equation}
where $\mathcal{L}^\ast_{z_k}(\hat{\rho}_\mathrm{B}(t))= \frac{1}{2}[z_k\hat{\rho}_\mathrm{B}(t),z_k^\dagger]+\frac{1}{2}[z_k,\hat{\rho}_\mathrm{B}(t)z_k^\dagger]$, $k=1,2$.

\begin{figure}[ht]
\includegraphics[scale=0.24]{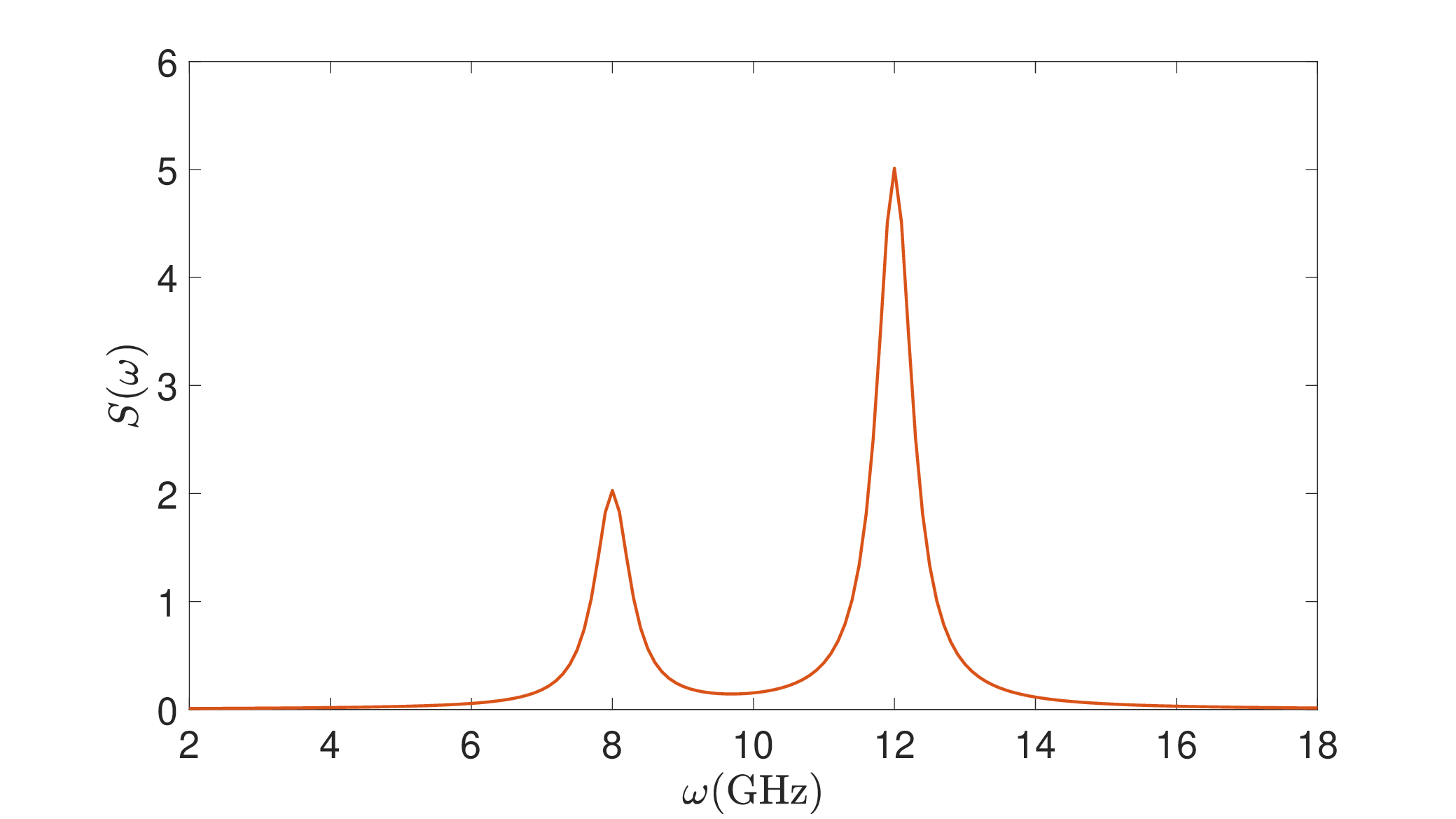}
\caption{\label{two_Lorentzian}The two-Lorentzian spectra.}
\end{figure}

\begin{figure}[ht]
\includegraphics[scale=0.25]{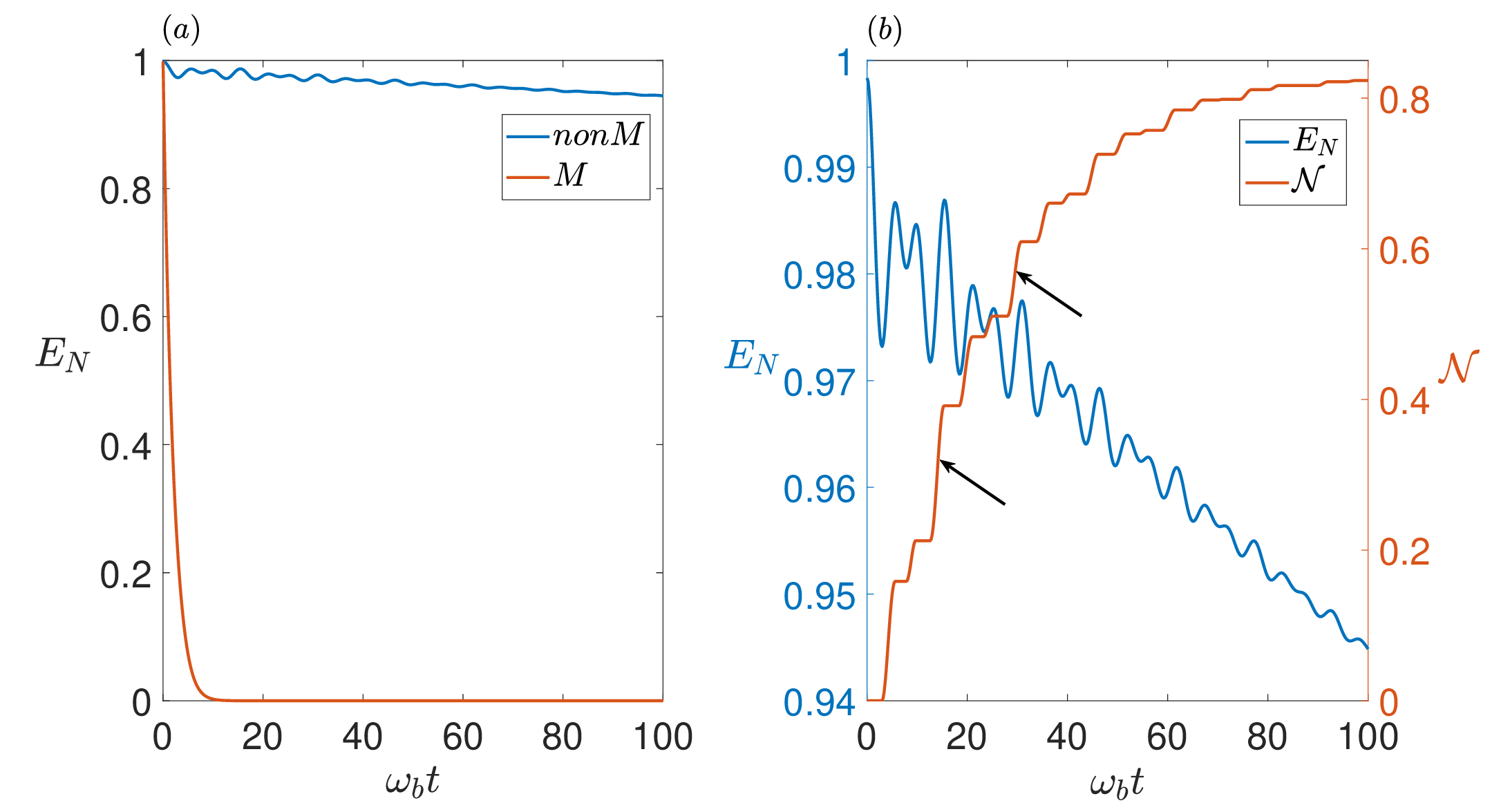}
\caption{\label{two_L_entanglement}($a$) Comparison of the evolution of the logarithmic negativity $E_N$ with the dimensionless time $\omega_bt$ in the non-Markovian quantum channel disturbed by two-Lorentzian noise and the corresponding Markovian quantum channel. ($b$) (Color online) The logarithmic negativity $E_N$ (blue full curve) and the BLP quantifier $\mathcal{N}$ (red full curve) as functions of the dimensionless time $\omega_bt$ in the non-Markovian quantum channel disturbed by two-Lorentzian noise.}
\end{figure}

\begin{figure}[ht]
\includegraphics[scale=0.25]{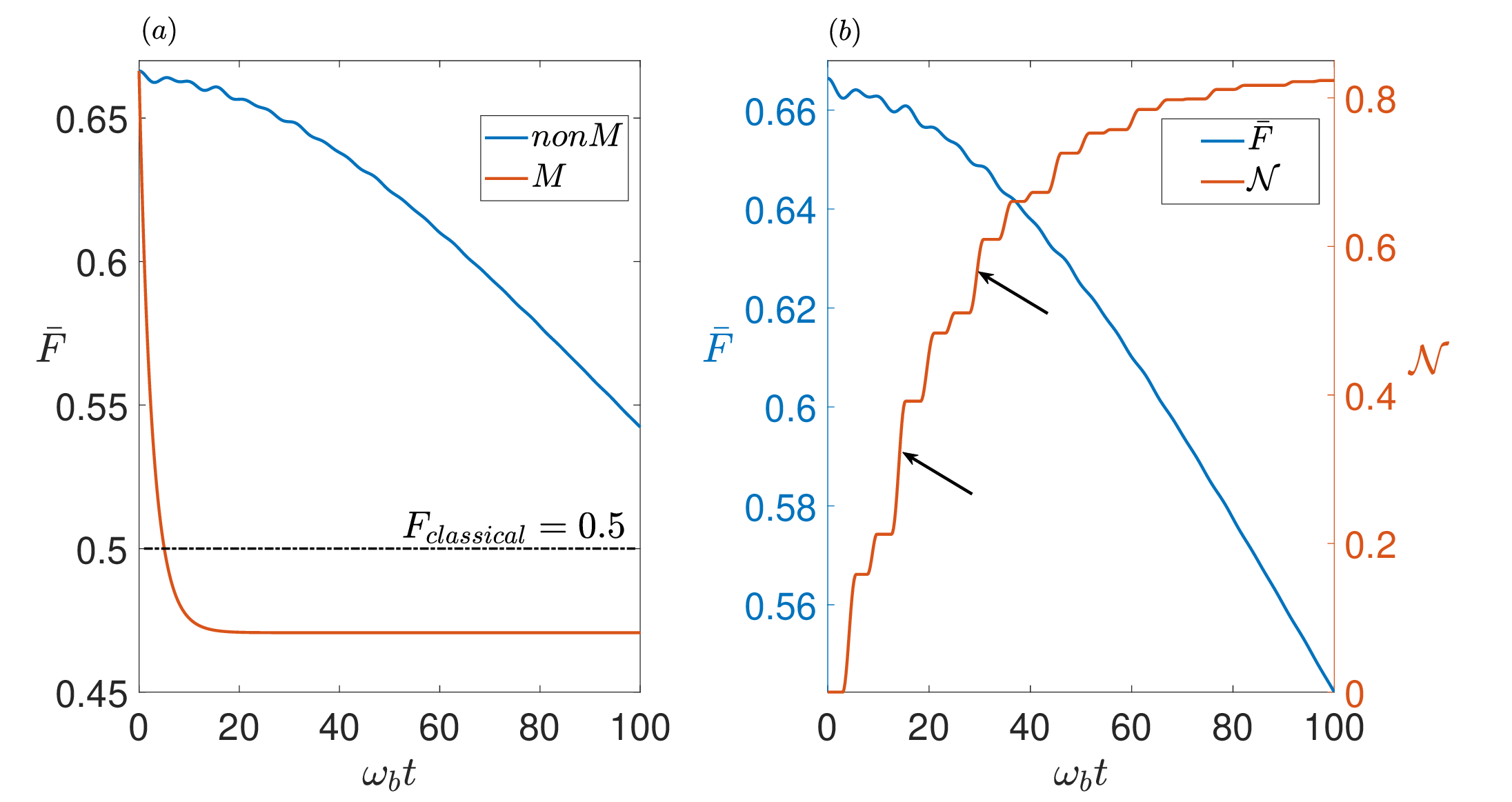}
\caption{\label{two_L_fidelity}($a$) Comparison of the evolution of the average fidelity $\bar{F}$ with the dimensionless time $\omega_bt$ in the non-Markovian quantum channel disturbed by two-Lorentzian noise and the corresponding Markovian quantum channel. The horizontal dash-dotted line represents the upper boundary of classical teleportation, $F_{classical}=0.5$. ($b$) (Color online) The average fidelity $\bar{F}$ (blue full curve) and the BLP quantifier $\mathcal{N}$ (red full curve) as functions of the dimensionless time $\omega_bt$ in the non-Markovian quantum channel disturbed by two-Lorentzian noise.}
\end{figure}

The corresponding
parameters are chosen as $\omega_1=8\ \rm{GHz}$, $\omega_2=12\ \mathrm{GH}z$, $\gamma_1=0.6\ \mathrm{GH}z$, $\gamma_2=0.6\ \mathrm{GH}z$, $\kappa_1=2\ \mathrm{GH}z$ and $\kappa_2=5\ \mathrm{GH}z$. Its spectrum is shown in Fig.~\ref{two_Lorentzian}. Different from Lorentzian noise, the spectrum of the two-Lorentzian noise has two peaks. The effects of non-Markovian quantum channels on the entangled state and CV quantum teleportation are investigated using the two-mode squeezed vacuum state with the squeezing parameter $r=0.346\ (-3\ \mathrm{dB})$ as the entangled state and a coherent state as the input state. The results are presented in Fig.~\ref{two_L_entanglement} and Fig.~\ref{two_L_fidelity}, respectively. Fig.~\ref{two_L_entanglement}($a$) demonstrates the logarithmic negativity of the entangled state can be maintained by the non-Markovian dynamics of the quantum channel compared with the corresponding Markovian quantum channel. The increase of non-Markovianity of the quantum channel can bring about the recovery of the entanglement as clearly illustrated in Fig.~\ref{two_L_entanglement}($b$). Similarly, Fig.~\ref{two_L_fidelity}($a$) shows the decay of the average fidelity of CV quantum teleportation can be effectively suppressed under the influence of the non-Markovian dynamics of the quantum channel disturbed by two-Lorentzian noise, compared with the corresponding Markovian quantum channel. The fidelity $\bar{F}> F_{classical}$ indicates that the entanglement remains used as the quantum resource in the case of a non-Markovian quantum channel disturbed by two-Lorentzian noise. Fig.~\ref{two_L_fidelity}($b$) shows the increase of non-Markovianity of the quantum channel can likewise lead to the recovery of $\bar{F}$.
Moreover, in Fig.~\ref{two_L_entanglement}($b$) and Fig.~\ref{two_L_fidelity}($b$), the larger increases in non-Markovianity are interspersed with smaller increases, as indicated by the arrows, unlike quantum channels disturbed by Lorentzian noise, where significant increases in non-Markovianity appear only at the beginning of the transmission. Meanwhile, corresponding to the larger increase in non-Markovianity, the recovery of both entanglement and the average fidelity exhibit greater magnitude. This also demonstrates that the strength of the effect is determined by the magnitude of each increase in the non-Markovianity of the quantum channel.
These results are consistent with the case of Lorentzian noise.

\begin{figure*}
\subfigure[Squeezed states\qquad \qquad]{\includegraphics[scale=0.3]{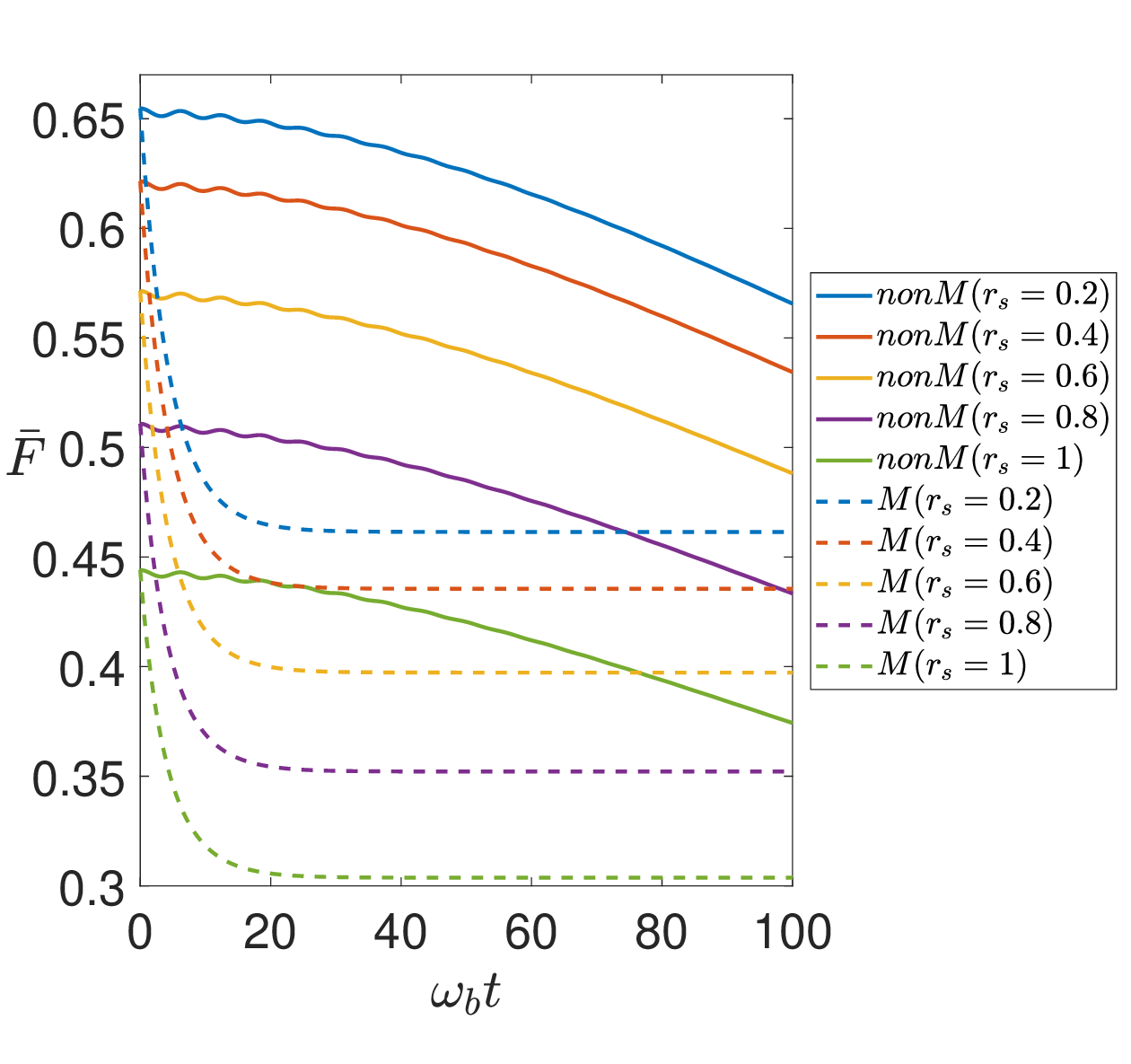}}
\subfigure[Cat states\qquad \quad ]{\includegraphics[scale=0.3]{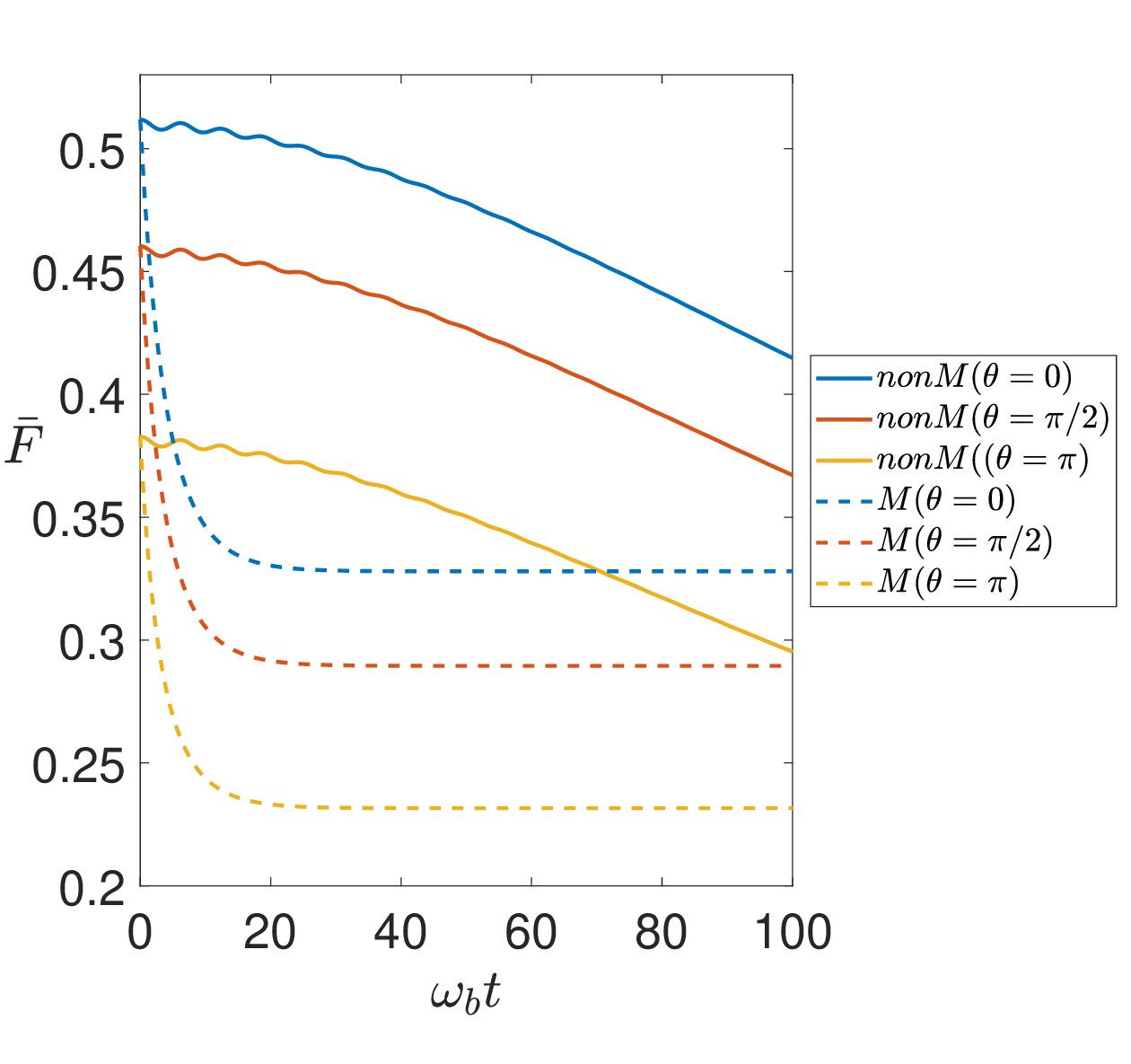}}
\caption{\label{other_input}($a$) $\bar{F}$ for the squeezed states with different squeezing parameters through the non-Markovian quantum channel and the corresponding Markovian quantum channel. $|\alpha_s|=1$. ($b$) $\bar{F}$ for the cat states with different phases through the non-Markovian quantum channel and the corresponding Markovian quantum channel. $|\alpha_c|=1$. }
\end{figure*}

\begin{figure*}
\centering
\includegraphics[scale=0.5]{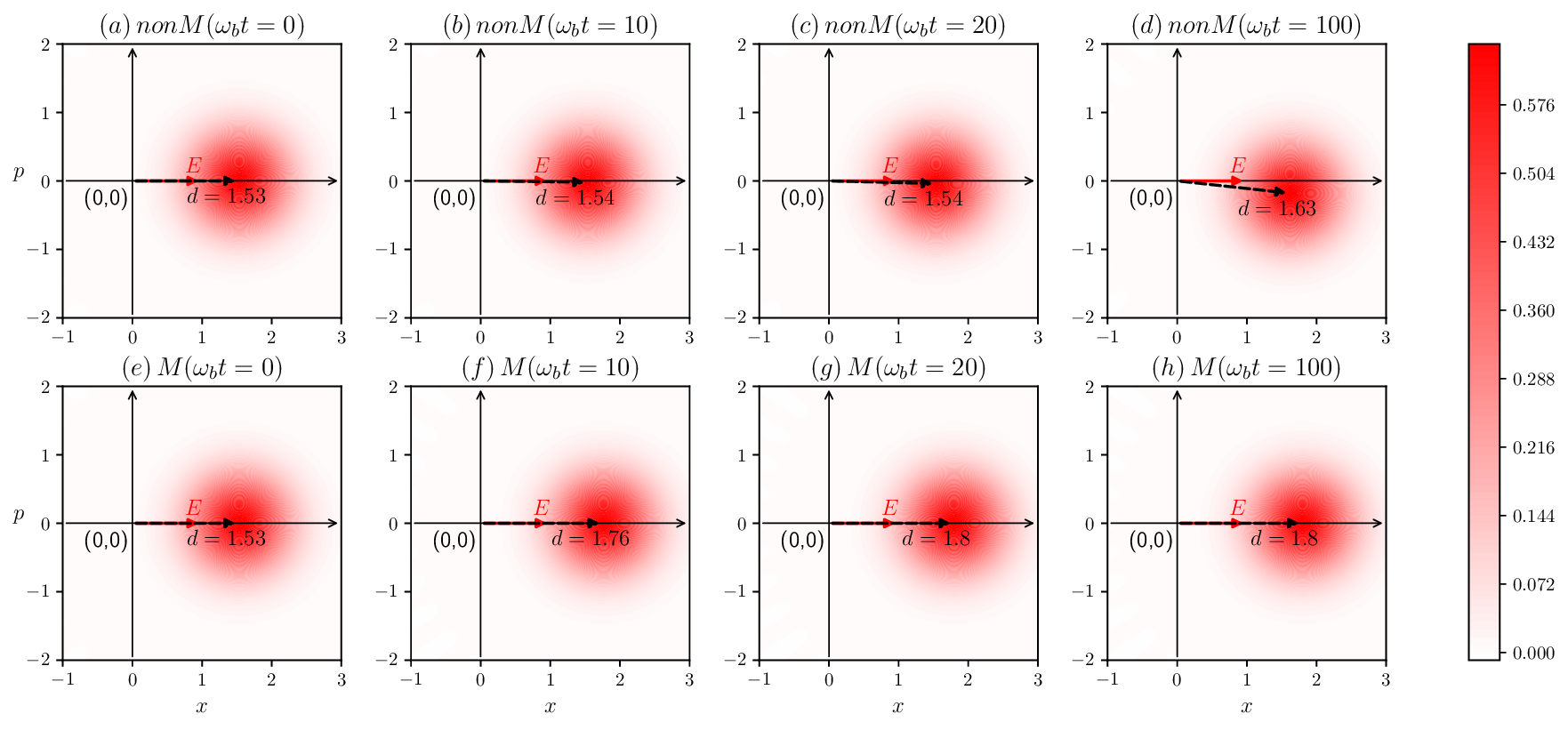}
\caption{\label{wigner_coherent} Wigner functions of the output state $|\alpha\rangle_{out}$ with different transfer times. Subplots $(a)$ - $(d)$ describe the evolution in the non-Markovian quantum channel. Subplots $(e)$ - $(h)$ describe the evolution in the corresponding Markovian quantum channel. In each subplot, the intersection of the two black solid lines with arrows is denoted as $(0,0)$, indicating the origin of phase space. The red solid arrow with the tail noted as $E$ indicates the amplitude of the input state $|\alpha\rangle$ at Alice, and the amplitude of the output state $|\alpha\rangle_{out}$ at Bob is represents by the black dashed arrow, with the exact amplitude $d$ marked below.}
\end{figure*}
\section{Teleportation of other states}
Besides coherent states, we also investigate the performance of quantum teleportation with other input states, such as squeezed states and cat states. Similarly, all these analyses are based on the non-Markovian quantum channel disturbed by Lorentzian noise with parameters $\gamma_0=0.8$ and $\kappa_0=4$.

The squeezed coherent state is written as $|\alpha\rangle_{s} = \hat{D}(\alpha_s)\hat{S}(\xi) |0\rangle$~\cite{PhysRevD.29.1107}, where
$\hat{S}(\xi) = \mathrm{exp}[\frac{1}{2}\xi^\ast \hat{a}^2-\frac{1}{2}\xi \hat{a}^{\dagger2}]$ is the squeezing operator with $\xi = r_s \mathrm{exp}(i\theta)$. Note that we distinguish the squeezing parameter $r_s$ from the squeezing parameter $r$ of the two-mode squeezed vacuum state. Also,
the cat state is denoted as $|\alpha \rangle_c=N^{-1}(|\alpha_c\rangle+e^{i\theta_c}|-\alpha_c\rangle)$, where $N$ is the normalization constant $N=\sqrt{2(1+e^{-2|\alpha_{c}|^{2}}cos\theta_c)}$, $|\pm \alpha_c\rangle$ are coherent states with the complex number $\alpha_c$, and $\theta_c$ is the phase~\cite{agarwal2012quantum}. In addition, the cat states with $\theta_c=0$ and $\theta_c=\pi$ are called even cat state and odd cat state, respectively. Note that the teleportation fidelity is not less than $2/3$ is the prerequisite for the cat state to be successfully teleported in CV quantum teleportation~\cite{ban2004phase}, and thus the squeezing parameter $r$ of the two-mode squeezed vacuum state is set to $r=0.4$ $($i.e., $-3.47\ \mathrm{dB})$. Moreover, to fairly compare the difference in the effect of different input states, we set $\alpha_s=\alpha_c=\alpha$ and $|\alpha|=1$.


\begin{figure*}
\centering
\includegraphics[scale=0.5]{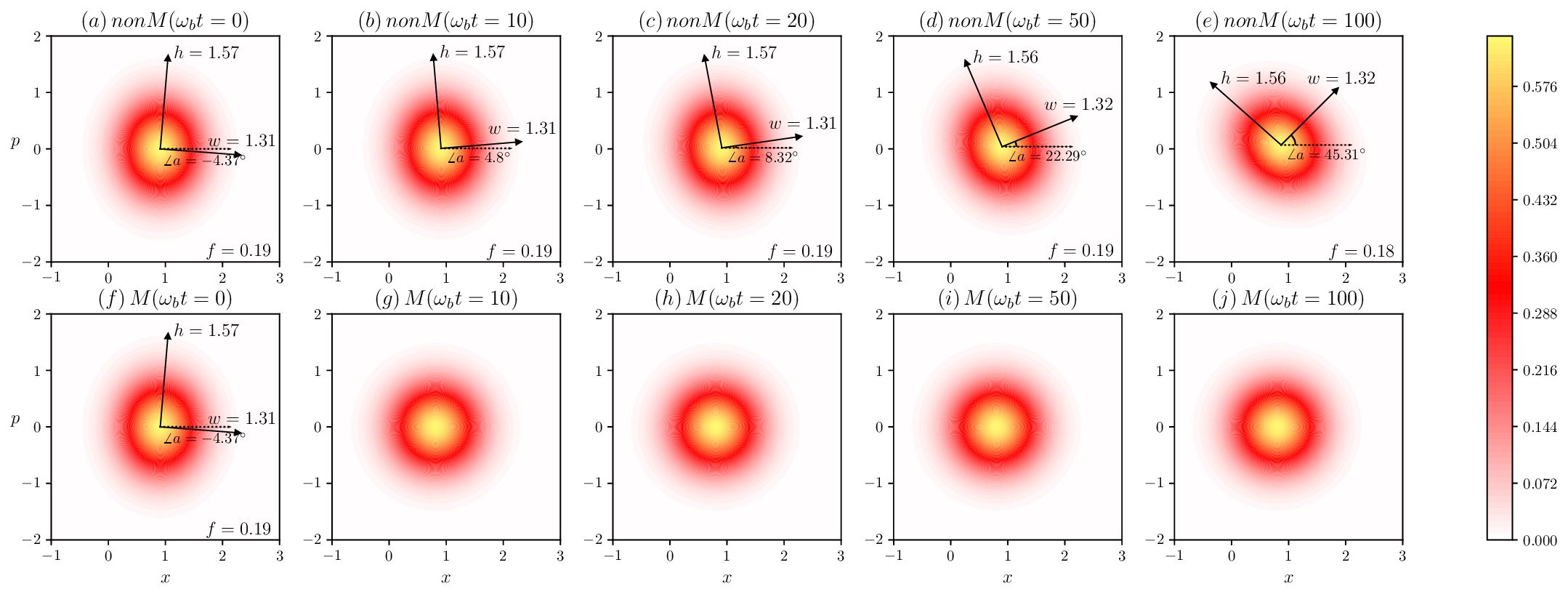}
\caption{\label{wigner_squeeze}
Wigner functions of the output states with different transmission time points of a squeezed state. Subplots ($a$) – ($e$) describe the evolution in the non-Markovian quantum channel. The major and minor semi-axis of the output state at Bob are depicted by solid black arrows labeled with specific values $h$ and $w$ at their tails, respectively. The phase shift of the output state in relation to the horizontal dashed line is indicated by $\angle a$. The precise angle is also calculated and labeled. The flattening of an ellipse is denoted by $f=(h-w)/w$. This parameter describes the degree of flattening of an ellipse and ranges between 0 and 1. The closer the value is to 1, the flatter the shape is. When $f=0$, it indicates a circular shape. Subplots ($f$) - ($j$) describe the evolution in the Markovian quantum channel.
}
\end{figure*}

The average fidelity for the squeezed states with different squeezing parameters and the cat states with different phases through the non-Markovian quantum channel and the corresponding Markovian quantum channel are illustrated in Fig.~\ref{other_input}$(a)$ and Fig.~\ref{other_input}$(b)$, respectively. The effects of the non-Markovian quantum channel on the quantum teleportation for squeezed states and cat states are consistent with the case of coherent states as shown in the Fig.~\ref{fidelity_parameter_r_k}. Under the influence of non-Markovian quantum channels, the average fidelity $\bar{F}$ exhibits fluctuating recoveries and holds relatively stable during the early stage of teleportation, e.g., $0 \leq \omega_bt \leq 20$, and then decays with the increase of the transmission time. While for the corresponding Markovian quantum channels, $\bar{F}$ decays faster from the beginning of transmission.

To clearly illustrate the effect of the non-Markovian quantum channel on the performance of quantum teleportation, Wigner functions of the reconstructed states at Bob with different input states are investigated.
When a coherent state is teleported, the Wigner functions of the output state $|\alpha\rangle_{out}$ at Bob at different times are plotted on the top $(a)$ - $(d)$ of Fig.~(\ref{wigner_coherent}), whilst the Wigner functions of $|\alpha\rangle_{out}$ in the corresponding Markovian quantum channel at the same times are plotted on the bottom $(e)$ - $(h)$. Both in the non-Markovian and Markovian quantum channels, the Wigner function of the output state $|\alpha\rangle_{out}$ gradually deviates from its initial position as the transmission time increases. The amplitude of the output state evolves from $d=1.53$ to $d=1.63$ shown in subplot $(d)$ and $d=1.80$ shown in subplot $(h)$ in the non-Markovian quantum channel and the corresponding Markovian quantum channel, respectively. However, for any same transmission time, the Wigner function of $|\alpha\rangle_{out}$ through the non-Markovian quantum channel is closer to the input state $|\alpha\rangle$ than that one through the corresponding Markovian quantum channel.
This demonstrates less information loss and better performance in the non-Markovian quantum channel. Especially, when $0\leq \omega_b t \leq20$, compared with the fast change of the Wigner functions of $|\alpha\rangle_{out}$ through the corresponding Markovian quantum channel described in subplots $(e)$ - $(g)$, the Wigner functions of $|\alpha\rangle_{out}$ through the non-Markovian quantum channel described in subplots $(a)$ - $(c)$ are stable and almost the same as the Wigner function described in subplot $(a)$ for the theoretical output state at $w_b t=0$.

\begin{figure*}
\centering
\includegraphics[scale=0.4]{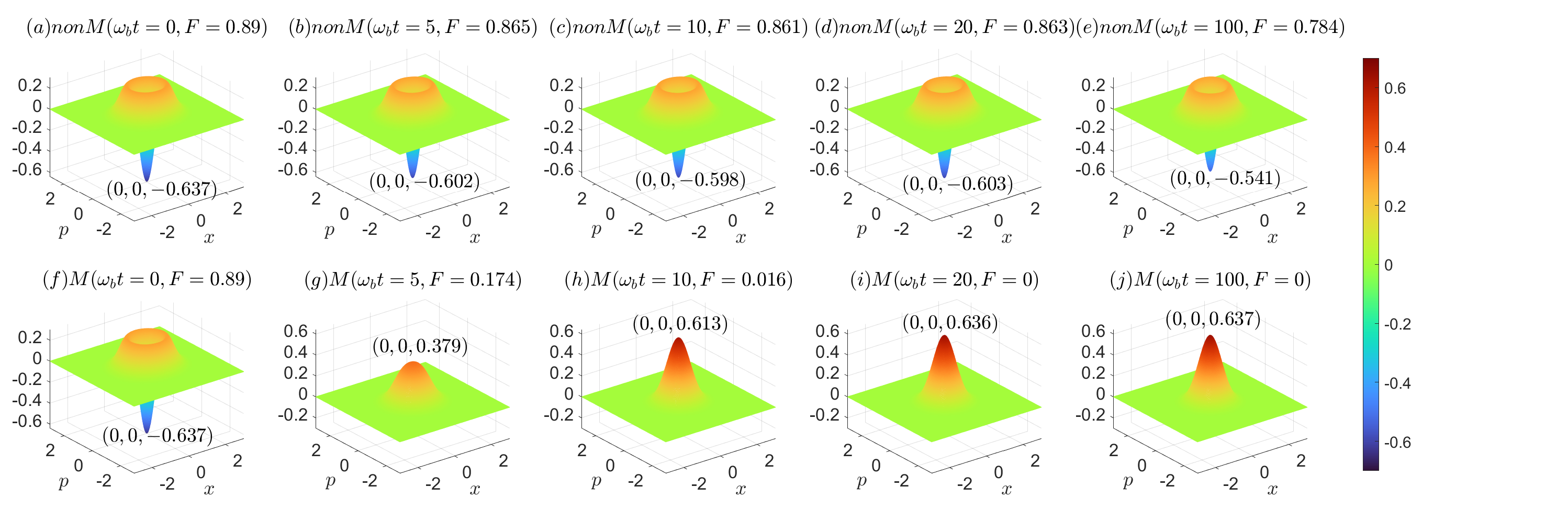}
\caption{\label{wigner_cat} Wigner functions for the output states with different transmission time points of a cat state. The non-classicality of the cat state can be maintained in the non-Markovian quantum channel.}
\end{figure*}

In the case of the squeezed state $|\alpha\rangle_{s}$ teleportation with a squeezing parameter $r_s=1$, the Wigner functions for the output states with different transmission times through a non-Markovian quantum channel and the corresponding Markovian quantum channel are plotted on the top and bottom rows of Fig.~\ref{wigner_squeeze}, respectively.
Fig.~\ref{wigner_squeeze}$(a)$ - $(e)$ illustrate the evolution of the Wigner function of the output state with the increase of the transmission time in the non-Markovian quantum channel. The flattening of an ellipse, denoted as $f$, remains around 0.19. This indicates that all of these Wigner functions maintain a compressed shape. However, phase shift of the output state is observed. As the transmission time increases, the angle of rotation experiences an increment from $-{4.37}^\circ$ to ${45.31}^\circ$.
While through the corresponding Markovian quantum channel, the compressed shape observed in the Wigner function of the output state quickly evolves into a circular shape, as shown in subplots $(f)$ - $(j)$ in Fig.~\ref{wigner_squeeze}. For the case of teleporting an odd cat state $|\alpha \rangle_c$ with $\theta_c=\pi$, the Wigner functions are shown in Fig.~\ref{wigner_cat}. Fig.~\ref{wigner_cat}$(a)$ - $(e)$ show the Wigner function with different transmission times through the non-Markovian quantum channel. In all these Wigner functions, the value at the origin $(0,0)$ is less than $0$, which indicates that the non-classicality can be maintained in the non-Markovian quantum channel. While in the corresponding Markovian quantum channel, the non-classicality disappears rapidly and the cat state degenerates to a Gaussian state as shown in Fig.~\ref{wigner_cat}$(f)$ - $(j)$. Similar to the case of coherent-state teleportation, the fidelities of the squeezed-state and cat-state teleportation decay slower in non-Markovian quantum channel than that the Markovian one such that the average fidelity can be kept at a high level.

\begin{figure}[h]
\includegraphics[scale=0.3]{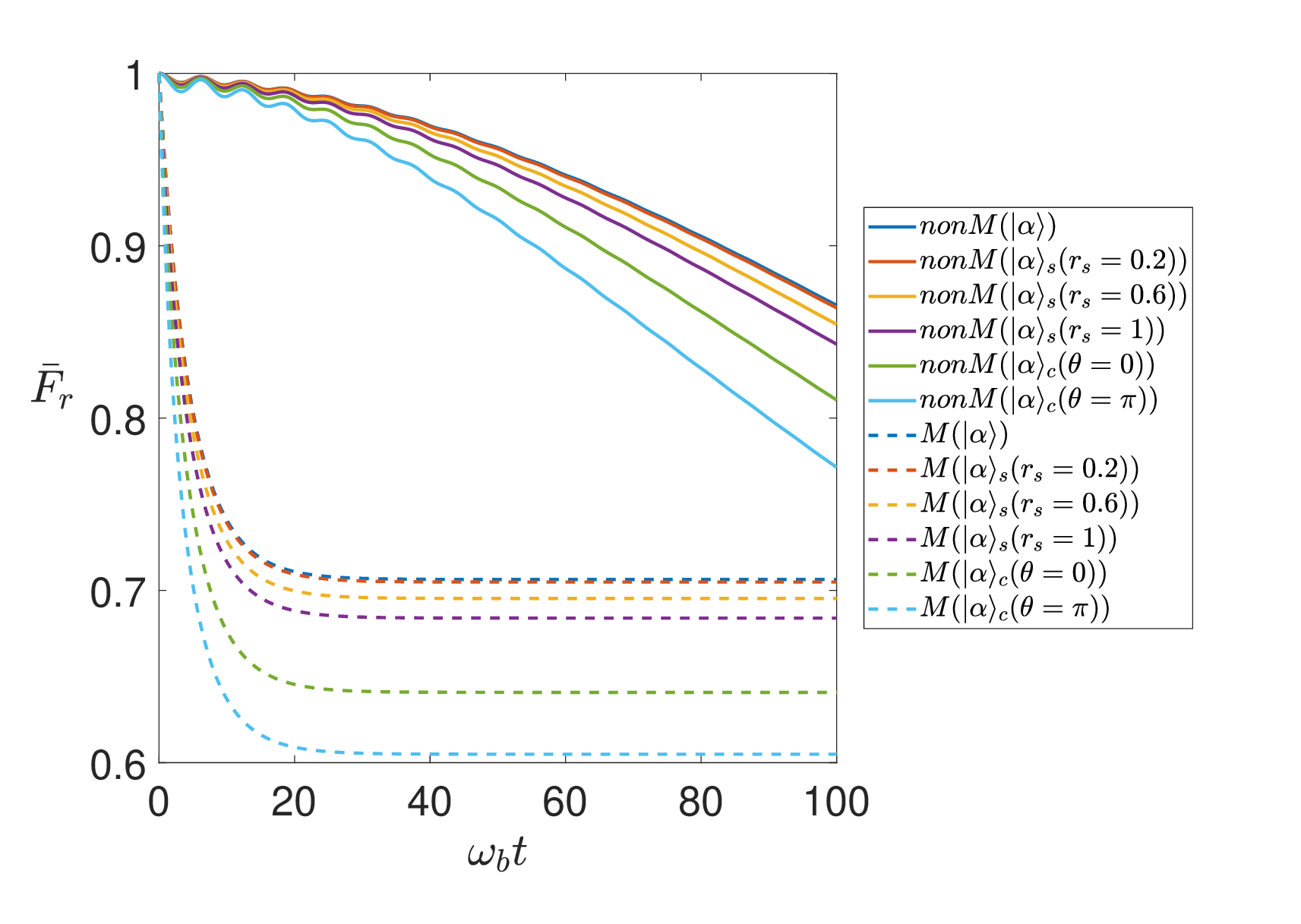}
\caption{\label{3_state_compare}The relative average fidelity $\bar{F}_r$ varying with different input states. The same parameters $\alpha$ of them are set $|\alpha|=1$.}
\end{figure}

In addition, there are inherent differences in the efficiency of information transmission for quantum teleportation with different input states, as shown in Fig.~\ref{other_input}. To avoid any errors resulting from these unavoidable differences and evaluate the performance of the different input states teleportation through the non-Markovian quantum channel, the concept of the relative average fidelity is introduced. It serves as a numerical indicator that quantifies the relative change in the average fidelity with respect to the average fidelity of the theoretical state due to the non-Markovian dynamics of the quantum channels. The relative average fidelity is defined as,
\begin{equation}
\bar{F}_r(\rho_{q(\mathrm{R},\mathrm{B})},t)=\frac{\bar{F}(\rho_{q(\mathrm{R},\mathrm{B})},t)}{\bar{F}(\rho_{q(\mathrm{R},\mathrm{B})},0)},
\end{equation}
where $t$ is the transit time, $\bar{F}(\rho_{q(\mathrm{R},\mathrm{B})},t)$ is calculated based on Eq.~(\ref{f_av}). Then the relative average fidelities for squeeze states with different $r_s$, cat states with different $\theta_c$ (the even cat state and the odd cat state), and the coherent state are calculated. The results are shown in Fig.~\ref{3_state_compare}. Meanwhile, the same indices are calculated for the case of the corresponding Markovian quantum channel and plotted in the same figure.

Fig.~\ref{3_state_compare} reveals the relative average fidelities for different input states can remain relatively stable for a certain transmission time under the influence of the non-Markovian dynamics of the quantum channel.
Concretely, when $\omega_b t \leq 20$, all the relative average fidelities fluctuate near unit 1, where the minimum value is $F_r=0.9791$.
However, in the same environment, the performance of different input states is different for both non-Markovian and Markovian. The coherent state $|\alpha\rangle$ is the most robust, followed by the squeezed coherent state $|\alpha \rangle_s$, and the cat state $|\alpha\rangle_c$ is the worst, especially the odd cat state $|\alpha\rangle_c(\theta=\pi)$. This is consistent with the fact that the cat state is more susceptible to decoherence caused by the disturbance of the environment.

\section{Discussion}
\subsection{Experimental feasibility}
The physical realization of an augmented system has been investigated, and it can be achieved by two coupled cavities~\cite{xue2019modeling,wiseman1994all}. One of the cavities, driven by quantum white noise, serves as the ancillary system to realize the internal modes of the non-Markovian environment. Through an optical crystal, the optical mode in the ancillary cavity is directly coupled to the optical mode in the other cavity; i.e., the principal cavity. Similarly, in the experiment of continuous teleportation, the non-Markovian quantum channel can be realized by optical fiber which couples to a waveguide cavity through their coupler~\cite{groblacher2013highly}. In this way, the mode in the fiber can be disturbed by quantum Lorentzian noise, which would exhibit non-Markovian behavior. To realize quantum colored noise with a complicated spectrum, a network of waveguide cavities is required to serve as the ancillary system where one of them is driven by quantum white noise such that a fictitious output of one cavity takes the quantum colored noise~\cite{xue2019modeling}. In an experiment, we can also couple the fictitious output to an optical fiber; i.e., the non-Markovian quantum channel, through a carefully designed coupler with sophisticated hole structures~\cite{groblacher2013highly}.

However, several potential challenges should be taken into consideration. On the one hand, to generate a given quantum colored noise with a complicated spectrum, the parameters of every cavity in the network for the ancillary system should be locked to keep the shape of the spectrum to be stable in an experiment. The parameters can be the frequency of each cavity, the damping rate of each cavity to a field or the coupling strength between two cavities~\cite{xue2019modeling}. Although some techniques are valid, for example, frequency locking of a cavity~\cite{cygan2013spectral}, locking multiple parameters simultaneously would not be easy in an experiment. To tackle the locking challenge, model reduction methods~\cite{zhou1996robust} can be considered to obtain an effective network with less number of cavities for the ancillary system so as to relieve the heavy task of parameter locking in an experiment. On the other hand, supposing the quantum channel is realized by an optical fiber, how to design the coupler with sufficient coupling strength between the channel and the fictitious output is also an important issue since the coupling strength can affect the amplitude of the noise. One may carefully design the hole structure for the coupler or use special materials for the hole such that the coupling strength can be sufficient.
%
%
%
\subsection{Comparison with other techniques}
Decoherence caused by noisy quantum channels is a significant issue in quantum information processing. Various methods have been investigated to mitigate this problem, including photon-subtracted two-mode squeezed vacuum state (PSTMSV)~\cite{kitagawa2006entanglement}, quantum scissors~\cite{xu2018improvement}, and quantum catalysis~\cite{hu2017continuous}. Among these methods, PSTMSV has great advantages in performance and experimental realization. PSTMSV is a method to obtain the high quality of entangled states by subtracting single photons from a two-mode squeezed vacuum state~\cite{kitagawa2006entanglement}. By directly improving the quality of entangled states, PSTMSV can enhance the performance of quantum teleportation.

However, it is important to note that quantum teleportation based on PSTMSV still suffers from the noise and decoherence in the quantum channel. As a consequence, PSTMSV imposes strict requirements on the squeezing aspect of the two-mode squeezed vacuum state for better information transmission performance. In contrast, our approach, non-Markovian quantum channels, aims to mitigate the decoherence from the perspective of quantum channels, which offers benefits for quantum teleportation based on any entangled states. Quantum teleportation with higher fidelity is available even if the entanglement of quantum resources is not quite high. Therefore, our approach can be combined with PSTMSV to further enhance the performance of quantum teleportation. This integration offers the opportunity to leverage the advantages of both techniques, resulting in an overall improvement in performance.

\section{Conclusion}
In this paper, we have constructed a non-Markovian quantum channel modeled by an augmented system and investigated the effect of the non-Markovian dynamics of the quantum channel on the performance of the CV quantum teleportation. Compared with the corresponding Markovian quantum channels, the non-Markovian dynamics of the quantum channels can effectively slow the degradation of the entanglement of the two-mode squeezed vacuum state, thus mitigating the deterioration of the performance of the CV quantum teleportation and maintaining the average fidelity relatively stable for a certain transmission time. The intensity of effects on both the entangled state and the CV quantum teleportation is closely related to the amount of information backflow per time. These results also work for the case of teleporting different input states, where coherent states have the best robustness.

Quantum teleportation is a crucial foundation of quantum communication networks. The robust quantum teleportation we proposed can mitigate the decoherence of quantum resources and achieve a higher fidelity in long distance information transmission. This advantage can effectively improve the performance and distance of information transmission, making it particularly valuable in quantum networks. In particular, for quantum networks with star-topology structures~\cite{sun2016quantum}, the robust quantum teleportation can extend the distance between the relay nodes and the central node, thereby expanding the coverage area for communication. Moreover, the robust quantum teleportation enables the communication of multiple pairs over long distances using fewer nodes in quantum teleportation networks~\cite{yonezawa2004demonstration}, which offers advantages in terms of cost-effectiveness and security.


\section*{ACKNOWLEDGMENT}
This work is supported by the National Natural Science Foundation of China (NSFC) under Grants No. 62273226, No. 61873162, Grant No. 61903316, and Shenzhen Fundamental Research Fund under the Grant No. JCYJ20190813165207290.

\appendix
\section{The solution to an augmented system model}
In this section, we review the solution to the master equation Eq.~(\ref{ms}) introduced in the main text, calculating the evolution of the density matrix of an augmented Markovian system.

Rewrite the master equation~(\ref{ms}) as
\begin{equation}
\dot{\hat{\rho}}(t)= -i[{\rm H}_p,\hat{\rho}(t)]-i[{\rm H}_a+{\rm H}_{pa},\hat{\rho}(t)]+\mathcal{L}^\ast_{L_a}(\hat{\rho}(t)).
\label{ms_2}
\end{equation}
In Eq.~(\ref{ms_2}), the first term can be represented with a superoperator,
\begin{equation}
\mathcal{L}_p \hat{\rho}(t)=-i[{\rm H}_p,\hat{\rho}(t)].
\end{equation}
The second term, which represents the internal dynamics of the ancillary system and the couplings to the principal system, can be described by another superoperator~\cite{xue2021gradient},
\begin{equation}
\mathcal{L}_a \hat{\rho}(t)= -i[{\rm H}_a+{\rm H}_{pa},\hat{\rho}(t)] .
\end{equation}
The last remaining term can be expressed with superoperators as well. In particular, it is a Lindblad superoperator,
\begin{equation}
\mathcal{L}^\ast_{L_a}(\hat{\rho}(t)) = \mathcal{D}_L \hat{\rho}(t).
\end{equation}
Therefore, the master equation~(\ref{ms_2}) can be expressed by the above three superoperators, namely
\begin{equation}
\begin{aligned}
\dot{\hat{\rho}}(t) &= (\mathcal{L}_p+\mathcal{L}_a+\mathcal{D}_L) \hat{\rho}(t)\\
&= \mathcal{L}\hat{\rho}(t).
\end{aligned}
\label{ms_l}
\end{equation}
The solution to Eq.~(\ref{ms_l}) is generally expressed in the following form,
\begin{equation}
\hat{\rho}(t) = \mathrm{exp}\{\mathcal{L}(t-t_0)\} \hat{\rho}(t_0),
\label{solution}
\end{equation}
where $t_0$ can be any point in the evolution, and generally $t_0 \leq t$. Substituting density matrix at the initial time $t_0=0$, $\hat{\rho}(0)$ into Eq.~(\ref{solution}), the density matrix at any time $t$ can be calculated. However, it is difficult to obtain an analytical expression of Eq.~(\ref{solution}) due to the calculations of the exponential of the superoperator $\mathcal{L}$.

Considering that the superoperator $\mathcal{L}$ is time-independent, the density matrix $\hat{\rho}(t)$ can be calculated by discretizing the evolution of an augment quantum system. Define the total evolution time of the initial density matrix $\hat{\rho}(0)$ in the augmented system as $T$.
Dividing the time $T$ into $N$ equal intervals, each time period can be expressed as $\Delta t=T/N$. $t_n$ denotes the $n$-th time $t_n = n \Delta t$ with $n=1, \dots, N$. Then, the density matrix $\hat{\rho}(t_n)$ can be calculated as
\begin{equation}
\hat{\rho}(t_n) = \mathcal{M}_n \dots \mathcal{M}_2 \mathcal{M}_1 \hat{\rho}(0), \qquad n=1, \dots, N.
\label{cal}
\end{equation}
In Eq.~(\ref{cal}) above, $\mathcal{M}_n$ is a matrix exponential form of the discretized superoperator of $\mathcal{L}$ at the time $t_n$, which can be calculated by
\begin{equation}
\mathcal{M}_n = \mathrm{exp}(\Delta t \mathcal{L}), \qquad n=1, \dots, N.
\end{equation}
Based on this method, the evolution of the initial density operator $\hat{\rho}(0)$ at the time $t_n, n=1, \dots, N$, in the augmented system can be calculated.

\nocite{*}


\begin{thebibliography}{42}%
\makeatletter
\providecommand \@ifxundefined [1]{%
 \@ifx{#1\undefined}
}%
\providecommand \@ifnum [1]{%
 \ifnum #1\expandafter \@firstoftwo
 \else \expandafter \@secondoftwo
 \fi
}%
\providecommand \@ifx [1]{%
 \ifx #1\expandafter \@firstoftwo
 \else \expandafter \@secondoftwo
 \fi
}%
\providecommand \natexlab [1]{#1}%
\providecommand \enquote  [1]{``#1''}%
\providecommand \bibnamefont  [1]{#1}%
\providecommand \bibfnamefont [1]{#1}%
\providecommand \citenamefont [1]{#1}%
\providecommand \href@noop [0]{\@secondoftwo}%
\providecommand \href [0]{\begingroup \@sanitize@url \@href}%
\providecommand \@href[1]{\@@startlink{#1}\@@href}%
\providecommand \@@href[1]{\endgroup#1\@@endlink}%
\providecommand \@sanitize@url [0]{\catcode `\\12\catcode `\$12\catcode
  `\&12\catcode `\#12\catcode `\^12\catcode `\_12\catcode `\%12\relax}%
\providecommand \@@startlink[1]{}%
\providecommand \@@endlink[0]{}%
\providecommand \url  [0]{\begingroup\@sanitize@url \@url }%
\providecommand \@url [1]{\endgroup\@href {#1}{\urlprefix }}%
\providecommand \urlprefix  [0]{URL }%
\providecommand \Eprint [0]{\href }%
\providecommand \doibase [0]{https://doi.org/}%
\providecommand \selectlanguage [0]{\@gobble}%
\providecommand \bibinfo  [0]{\@secondoftwo}%
\providecommand \bibfield  [0]{\@secondoftwo}%
\providecommand \translation [1]{[#1]}%
\providecommand \BibitemOpen [0]{}%
\providecommand \bibitemStop [0]{}%
\providecommand \bibitemNoStop [0]{.\EOS\space}%
\providecommand \EOS [0]{\spacefactor3000\relax}%
\providecommand \BibitemShut  [1]{\csname bibitem#1\endcsname}%
\let\auto@bib@innerbib\@empty
\bibitem [{\citenamefont {Bennett}\ \emph {et~al.}(1993)\citenamefont
  {Bennett}, \citenamefont {Brassard}, \citenamefont {Cr{\'e}peau},
  \citenamefont {Jozsa}, \citenamefont {Peres},\ and\ \citenamefont
  {Wootters}}]{bennett1993teleporting}%
  \BibitemOpen
  \bibfield  {author} {\bibinfo {author} {\bibfnamefont {C.~H.}\ \bibnamefont
  {Bennett}}, \bibinfo {author} {\bibfnamefont {G.}~\bibnamefont {Brassard}},
  \bibinfo {author} {\bibfnamefont {C.}~\bibnamefont {Cr{\'e}peau}}, \bibinfo
  {author} {\bibfnamefont {R.}~\bibnamefont {Jozsa}}, \bibinfo {author}
  {\bibfnamefont {A.}~\bibnamefont {Peres}},\ and\ \bibinfo {author}
  {\bibfnamefont {W.~K.}\ \bibnamefont {Wootters}},\ }\bibfield  {title}
  {\bibinfo {title} {Teleporting an unknown quantum state via dual classical
  and {E}instein-{P}odolsky-{R}osen channels},\ }\href@noop {} {\bibfield
  {journal} {\bibinfo  {journal} {Physical Review Letters}\ }\textbf {\bibinfo
  {volume} {70}},\ \bibinfo {pages} {1895} (\bibinfo {year}
  {1993})}\BibitemShut {NoStop}%
\bibitem [{\citenamefont {Bouwmeester}\ \emph {et~al.}(1997)\citenamefont
  {Bouwmeester}, \citenamefont {Pan}, \citenamefont {Mattle}, \citenamefont
  {Eibl}, \citenamefont {Weinfurter},\ and\ \citenamefont
  {Zeilinger}}]{bouwmeester1997experimental}%
  \BibitemOpen
  \bibfield  {author} {\bibinfo {author} {\bibfnamefont {D.}~\bibnamefont
  {Bouwmeester}}, \bibinfo {author} {\bibfnamefont {J.~W.}\ \bibnamefont
  {Pan}}, \bibinfo {author} {\bibfnamefont {K.}~\bibnamefont {Mattle}},
  \bibinfo {author} {\bibfnamefont {M.}~\bibnamefont {Eibl}}, \bibinfo {author}
  {\bibfnamefont {H.}~\bibnamefont {Weinfurter}},\ and\ \bibinfo {author}
  {\bibfnamefont {A.}~\bibnamefont {Zeilinger}},\ }\bibfield  {title} {\bibinfo
  {title} {Experimental quantum teleportation},\ }\href@noop {} {\bibfield
  {journal} {\bibinfo  {journal} {Nature}\ }\textbf {\bibinfo {volume} {390}},\
  \bibinfo {pages} {575} (\bibinfo {year} {1997})}\BibitemShut {NoStop}%
\bibitem [{\citenamefont {Aliferis}\ and\ \citenamefont
  {Leung}(2004)}]{aliferis2004computation}%
  \BibitemOpen
  \bibfield  {author} {\bibinfo {author} {\bibfnamefont {P.}~\bibnamefont
  {Aliferis}}\ and\ \bibinfo {author} {\bibfnamefont {D.~W.}\ \bibnamefont
  {Leung}},\ }\bibfield  {title} {\bibinfo {title} {Computation by
  measurements: a unifying picture},\ }\href@noop {} {\bibfield  {journal}
  {\bibinfo  {journal} {Physical Review A}\ }\textbf {\bibinfo {volume} {70}},\
  \bibinfo {pages} {062314} (\bibinfo {year} {2004})}\BibitemShut {NoStop}%
\bibitem [{\citenamefont {Ishizaka}\ and\ \citenamefont
  {Hiroshima}(2008)}]{ishizaka2008asymptotic}%
  \BibitemOpen
  \bibfield  {author} {\bibinfo {author} {\bibfnamefont {S.}~\bibnamefont
  {Ishizaka}}\ and\ \bibinfo {author} {\bibfnamefont {T.}~\bibnamefont
  {Hiroshima}},\ }\bibfield  {title} {\bibinfo {title} {Asymptotic
  teleportation scheme as a universal programmable quantum processor},\
  }\href@noop {} {\bibfield  {journal} {\bibinfo  {journal} {Physical Review
  Letters}\ }\textbf {\bibinfo {volume} {101}},\ \bibinfo {pages} {240501}
  (\bibinfo {year} {2008})}\BibitemShut {NoStop}%
\bibitem [{\citenamefont {Ishizaka}\ and\ \citenamefont
  {Hiroshima}(2009)}]{ishizaka2009quantum}%
  \BibitemOpen
  \bibfield  {author} {\bibinfo {author} {\bibfnamefont {S.}~\bibnamefont
  {Ishizaka}}\ and\ \bibinfo {author} {\bibfnamefont {T.}~\bibnamefont
  {Hiroshima}},\ }\bibfield  {title} {\bibinfo {title} {Quantum teleportation
  scheme by selecting one of multiple output ports},\ }\href@noop {} {\bibfield
   {journal} {\bibinfo  {journal} {Physical Review A}\ }\textbf {\bibinfo
  {volume} {79}},\ \bibinfo {pages} {042306} (\bibinfo {year}
  {2009})}\BibitemShut {NoStop}%
\bibitem [{\citenamefont {Briegel}\ \emph {et~al.}(1998)\citenamefont
  {Briegel}, \citenamefont {D{\"u}r}, \citenamefont {Cirac},\ and\
  \citenamefont {Zoller}}]{briegel1998quantum}%
  \BibitemOpen
  \bibfield  {author} {\bibinfo {author} {\bibfnamefont {H.~J.}\ \bibnamefont
  {Briegel}}, \bibinfo {author} {\bibfnamefont {W.}~\bibnamefont {D{\"u}r}},
  \bibinfo {author} {\bibfnamefont {J.~I.}\ \bibnamefont {Cirac}},\ and\
  \bibinfo {author} {\bibfnamefont {P.}~\bibnamefont {Zoller}},\ }\bibfield
  {title} {\bibinfo {title} {Quantum repeaters: the role of imperfect local
  operations in quantum communication},\ }\href@noop {} {\bibfield  {journal}
  {\bibinfo  {journal} {Physical Review Letters}\ }\textbf {\bibinfo {volume}
  {81}},\ \bibinfo {pages} {5932} (\bibinfo {year} {1998})}\BibitemShut
  {NoStop}%
\bibitem [{\citenamefont {Vaidman}(1994)}]{vaidman1994teleportation}%
  \BibitemOpen
  \bibfield  {author} {\bibinfo {author} {\bibfnamefont {L.}~\bibnamefont
  {Vaidman}},\ }\bibfield  {title} {\bibinfo {title} {Teleportation of quantum
  states},\ }\href@noop {} {\bibfield  {journal} {\bibinfo  {journal} {Physical
  Review A}\ }\textbf {\bibinfo {volume} {49}},\ \bibinfo {pages} {1473}
  (\bibinfo {year} {1994})}\BibitemShut {NoStop}%
\bibitem [{\citenamefont {Braunstein}\ and\ \citenamefont
  {Kimble}(1998)}]{braunstein1998teleportation}%
  \BibitemOpen
  \bibfield  {author} {\bibinfo {author} {\bibfnamefont {S.~L.}\ \bibnamefont
  {Braunstein}}\ and\ \bibinfo {author} {\bibfnamefont {H.~J.}\ \bibnamefont
  {Kimble}},\ }\bibfield  {title} {\bibinfo {title} {Teleportation of
  continuous quantum variables},\ }\href@noop {} {\bibfield  {journal}
  {\bibinfo  {journal} {Physical Review Letters}\ }\textbf {\bibinfo {volume}
  {80}},\ \bibinfo {pages} {869} (\bibinfo {year} {1998})}\BibitemShut
  {NoStop}%
\bibitem [{\citenamefont {Furusawa}\ \emph {et~al.}(1998)\citenamefont
  {Furusawa}, \citenamefont {S{\o}rensen}, \citenamefont {Braunstein},
  \citenamefont {Fuchs}, \citenamefont {Kimble},\ and\ \citenamefont
  {Polzik}}]{furusawa1998unconditional}%
  \BibitemOpen
  \bibfield  {author} {\bibinfo {author} {\bibfnamefont {A.}~\bibnamefont
  {Furusawa}}, \bibinfo {author} {\bibfnamefont {J.~L.}\ \bibnamefont
  {S{\o}rensen}}, \bibinfo {author} {\bibfnamefont {S.~L.}\ \bibnamefont
  {Braunstein}}, \bibinfo {author} {\bibfnamefont {C.~A.}\ \bibnamefont
  {Fuchs}}, \bibinfo {author} {\bibfnamefont {H.~J.}\ \bibnamefont {Kimble}},\
  and\ \bibinfo {author} {\bibfnamefont {E.~S.}\ \bibnamefont {Polzik}},\
  }\bibfield  {title} {\bibinfo {title} {Unconditional quantum teleportation},\
  }\href@noop {} {\bibfield  {journal} {\bibinfo  {journal} {Science}\ }\textbf
  {\bibinfo {volume} {282}},\ \bibinfo {pages} {706} (\bibinfo {year}
  {1998})}\BibitemShut {NoStop}%
\bibitem [{\citenamefont {Takei}\ \emph {et~al.}(2005)\citenamefont {Takei},
  \citenamefont {Aoki}, \citenamefont {Koike}, \citenamefont {Yoshino},
  \citenamefont {Wakui}, \citenamefont {Yonezawa}, \citenamefont {Hiraoka},
  \citenamefont {Mizuno}, \citenamefont {Takeoka}, \citenamefont {Ban},\ and\
  \citenamefont {Furusawa}}]{takei2005experimental}%
  \BibitemOpen
  \bibfield  {author} {\bibinfo {author} {\bibfnamefont {N.}~\bibnamefont
  {Takei}}, \bibinfo {author} {\bibfnamefont {T.}~\bibnamefont {Aoki}},
  \bibinfo {author} {\bibfnamefont {S.}~\bibnamefont {Koike}}, \bibinfo
  {author} {\bibfnamefont {K.~I.}\ \bibnamefont {Yoshino}}, \bibinfo {author}
  {\bibfnamefont {K.}~\bibnamefont {Wakui}}, \bibinfo {author} {\bibfnamefont
  {H.}~\bibnamefont {Yonezawa}}, \bibinfo {author} {\bibfnamefont
  {T.}~\bibnamefont {Hiraoka}}, \bibinfo {author} {\bibfnamefont
  {J.}~\bibnamefont {Mizuno}}, \bibinfo {author} {\bibfnamefont
  {M.}~\bibnamefont {Takeoka}}, \bibinfo {author} {\bibfnamefont
  {M.}~\bibnamefont {Ban}},\ and\ \bibinfo {author} {\bibfnamefont
  {A.}~\bibnamefont {Furusawa}},\ }\bibfield  {title} {\bibinfo {title}
  {Experimental demonstration of quantum teleportation of a squeezed state},\
  }\href {https://doi.org/10.1103/PhysRevA.72.042304} {\bibfield  {journal}
  {\bibinfo  {journal} {Physical Review A}\ }\textbf {\bibinfo {volume} {72}},\
  \bibinfo {pages} {042304} (\bibinfo {year} {2005})}\BibitemShut {NoStop}%
\bibitem [{\citenamefont {Lee}\ \emph {et~al.}(2011)\citenamefont {Lee},
  \citenamefont {Benichi}, \citenamefont {Takeno}, \citenamefont {Takeda},
  \citenamefont {Webb}, \citenamefont {Huntington},\ and\ \citenamefont
  {Furusawa}}]{lee2011teleportation}%
  \BibitemOpen
  \bibfield  {author} {\bibinfo {author} {\bibfnamefont {N.}~\bibnamefont
  {Lee}}, \bibinfo {author} {\bibfnamefont {H.}~\bibnamefont {Benichi}},
  \bibinfo {author} {\bibfnamefont {Y.}~\bibnamefont {Takeno}}, \bibinfo
  {author} {\bibfnamefont {S.}~\bibnamefont {Takeda}}, \bibinfo {author}
  {\bibfnamefont {J.}~\bibnamefont {Webb}}, \bibinfo {author} {\bibfnamefont
  {E.}~\bibnamefont {Huntington}},\ and\ \bibinfo {author} {\bibfnamefont
  {A.}~\bibnamefont {Furusawa}},\ }\bibfield  {title} {\bibinfo {title}
  {Teleportation of nonclassical wave packets of light},\ }\href@noop {}
  {\bibfield  {journal} {\bibinfo  {journal} {Science}\ }\textbf {\bibinfo
  {volume} {332}},\ \bibinfo {pages} {330} (\bibinfo {year}
  {2011})}\BibitemShut {NoStop}%
\bibitem [{\citenamefont {Fukui}\ and\ \citenamefont
  {Takeda}(2022)}]{fukui2022building}%
  \BibitemOpen
  \bibfield  {author} {\bibinfo {author} {\bibfnamefont {K.}~\bibnamefont
  {Fukui}}\ and\ \bibinfo {author} {\bibfnamefont {S.}~\bibnamefont {Takeda}},\
  }\bibfield  {title} {\bibinfo {title} {Building a large-scale quantum
  computer with continuous-variable optical technologies},\ }\href@noop {}
  {\bibfield  {journal} {\bibinfo  {journal} {Journal of Physics B: Atomic,
  Molecular and Optical Physics}\ }\textbf {\bibinfo {volume} {55}},\ \bibinfo
  {pages} {012001} (\bibinfo {year} {2022})}\BibitemShut {NoStop}%
\bibitem [{\citenamefont {Hu}\ \emph {et~al.}(2021)\citenamefont {Hu},
  \citenamefont {Huang}, \citenamefont {Sheng}, \citenamefont {Zhou},
  \citenamefont {Liu}, \citenamefont {Guo}, \citenamefont {Zhang},
  \citenamefont {Xing}, \citenamefont {Huang}, \citenamefont {Li},\ and\
  \citenamefont {Guo}}]{hu2021long}%
  \BibitemOpen
  \bibfield  {author} {\bibinfo {author} {\bibfnamefont {X.~M.}\ \bibnamefont
  {Hu}}, \bibinfo {author} {\bibfnamefont {C.~X.}\ \bibnamefont {Huang}},
  \bibinfo {author} {\bibfnamefont {Y.~B.}\ \bibnamefont {Sheng}}, \bibinfo
  {author} {\bibfnamefont {L.}~\bibnamefont {Zhou}}, \bibinfo {author}
  {\bibfnamefont {B.~H.}\ \bibnamefont {Liu}}, \bibinfo {author} {\bibfnamefont
  {Y.}~\bibnamefont {Guo}}, \bibinfo {author} {\bibfnamefont {C.}~\bibnamefont
  {Zhang}}, \bibinfo {author} {\bibfnamefont {W.~B.}\ \bibnamefont {Xing}},
  \bibinfo {author} {\bibfnamefont {Y.~F.}\ \bibnamefont {Huang}}, \bibinfo
  {author} {\bibfnamefont {C.~F.}\ \bibnamefont {Li}},\ and\ \bibinfo {author}
  {\bibfnamefont {G.~C.}\ \bibnamefont {Guo}},\ }\bibfield  {title} {\bibinfo
  {title} {Long-distance entanglement purification for quantum communication},\
  }\href@noop {} {\bibfield  {journal} {\bibinfo  {journal} {Physical Review
  Letters}\ }\textbf {\bibinfo {volume} {126}},\ \bibinfo {pages} {010503}
  (\bibinfo {year} {2021})}\BibitemShut {NoStop}%
\bibitem [{\citenamefont {Riera-S{\`a}bat}\ \emph {et~al.}(2021)\citenamefont
  {Riera-S{\`a}bat}, \citenamefont {Sekatski}, \citenamefont {Pirker},\ and\
  \citenamefont {D{\"u}r}}]{riera2021entanglement}%
  \BibitemOpen
  \bibfield  {author} {\bibinfo {author} {\bibfnamefont {F.}~\bibnamefont
  {Riera-S{\`a}bat}}, \bibinfo {author} {\bibfnamefont {P.}~\bibnamefont
  {Sekatski}}, \bibinfo {author} {\bibfnamefont {A.}~\bibnamefont {Pirker}},\
  and\ \bibinfo {author} {\bibfnamefont {W.}~\bibnamefont {D{\"u}r}},\
  }\bibfield  {title} {\bibinfo {title} {Entanglement-assisted entanglement
  purification},\ }\href {https://doi.org/10.1103/PhysRevLett.127.040502}
  {\bibfield  {journal} {\bibinfo  {journal} {Physical Review Letters}\
  }\textbf {\bibinfo {volume} {127}},\ \bibinfo {pages} {040502} (\bibinfo
  {year} {2021})}\BibitemShut {NoStop}%
\bibitem [{\citenamefont {Yan}\ \emph {et~al.}(2023)\citenamefont {Yan},
  \citenamefont {Zhou}, \citenamefont {Zhong},\ and\ \citenamefont
  {Sheng}}]{yan2023advances}%
  \BibitemOpen
  \bibfield  {author} {\bibinfo {author} {\bibfnamefont {P.~S.}\ \bibnamefont
  {Yan}}, \bibinfo {author} {\bibfnamefont {L.}~\bibnamefont {Zhou}}, \bibinfo
  {author} {\bibfnamefont {W.}~\bibnamefont {Zhong}},\ and\ \bibinfo {author}
  {\bibfnamefont {Y.~B.}\ \bibnamefont {Sheng}},\ }\bibfield  {title} {\bibinfo
  {title} {Advances in quantum entanglement purification},\ }\href@noop {}
  {\bibfield  {journal} {\bibinfo  {journal} {Science China Physics, Mechanics
  \& Astronomy}\ }\textbf {\bibinfo {volume} {66}},\ \bibinfo {pages} {250301}
  (\bibinfo {year} {2023})}\BibitemShut {NoStop}%
\bibitem [{\citenamefont {Drummond}\ and\ \citenamefont
  {Corney}(2001)}]{drummond2001quantum}%
  \BibitemOpen
  \bibfield  {author} {\bibinfo {author} {\bibfnamefont {P.}~\bibnamefont
  {Drummond}}\ and\ \bibinfo {author} {\bibfnamefont {J.~F.}\ \bibnamefont
  {Corney}},\ }\bibfield  {title} {\bibinfo {title} {Quantum noise in optical
  fibers. {I}. {S}tochastic equations},\ }\href@noop {} {\bibfield  {journal}
  {\bibinfo  {journal} {Journal of the Optical Society of America B}\ }\textbf
  {\bibinfo {volume} {18}},\ \bibinfo {pages} {139} (\bibinfo {year}
  {2001})}\BibitemShut {NoStop}%
\bibitem [{\citenamefont {Walls}\ and\ \citenamefont
  {Milburn}(1994)}]{walls1994atomic}%
  \BibitemOpen
  \bibfield  {author} {\bibinfo {author} {\bibfnamefont {D.~F.}\ \bibnamefont
  {Walls}}\ and\ \bibinfo {author} {\bibfnamefont {G.~J.}\ \bibnamefont
  {Milburn}},\ }\bibinfo {title} {Atomic optics},\ in\ \href@noop {} {\emph
  {\bibinfo {booktitle} {Quantum Optics}}}\ (\bibinfo  {publisher} {Springer
  Berlin Heidelberg},\ \bibinfo {year} {1994})\ pp.\ \bibinfo {pages}
  {315--340}\BibitemShut {NoStop}%
\bibitem [{\citenamefont {Furusawa}\ and\ \citenamefont
  {Takei}(2007)}]{furusawa2007quantum}%
  \BibitemOpen
  \bibfield  {author} {\bibinfo {author} {\bibfnamefont {A.}~\bibnamefont
  {Furusawa}}\ and\ \bibinfo {author} {\bibfnamefont {N.}~\bibnamefont
  {Takei}},\ }\bibfield  {title} {\bibinfo {title} {Quantum teleportation for
  continuous variables and related quantum information processing},\
  }\href@noop {} {\bibfield  {journal} {\bibinfo  {journal} {Physics Reports}\
  }\textbf {\bibinfo {volume} {443}},\ \bibinfo {pages} {97} (\bibinfo {year}
  {2007})}\BibitemShut {NoStop}%
\bibitem [{\citenamefont {Hofmann}\ \emph {et~al.}(2000)\citenamefont
  {Hofmann}, \citenamefont {Ide}, \citenamefont {Kobayashi},\ and\
  \citenamefont {Furusawa}}]{hofmann2000fidelity}%
  \BibitemOpen
  \bibfield  {author} {\bibinfo {author} {\bibfnamefont {H.~F.}\ \bibnamefont
  {Hofmann}}, \bibinfo {author} {\bibfnamefont {T.}~\bibnamefont {Ide}},
  \bibinfo {author} {\bibfnamefont {T.}~\bibnamefont {Kobayashi}},\ and\
  \bibinfo {author} {\bibfnamefont {A.}~\bibnamefont {Furusawa}},\ }\bibfield
  {title} {\bibinfo {title} {Fidelity and information in the quantum
  teleportation of continuous variables},\ }\href@noop {} {\bibfield  {journal}
  {\bibinfo  {journal} {Physical Review A}\ }\textbf {\bibinfo {volume} {62}},\
  \bibinfo {pages} {062304} (\bibinfo {year} {2000})}\BibitemShut {NoStop}%
\bibitem [{\citenamefont {Vidal}\ and\ \citenamefont
  {Werner}(2002)}]{vidal2002computable}%
  \BibitemOpen
  \bibfield  {author} {\bibinfo {author} {\bibfnamefont {G.}~\bibnamefont
  {Vidal}}\ and\ \bibinfo {author} {\bibfnamefont {R.~F.}\ \bibnamefont
  {Werner}},\ }\bibfield  {title} {\bibinfo {title} {Computable measure of
  entanglement},\ }\href@noop {} {\bibfield  {journal} {\bibinfo  {journal}
  {Physical Review A}\ }\textbf {\bibinfo {volume} {65}},\ \bibinfo {pages}
  {032314} (\bibinfo {year} {2002})}\BibitemShut {NoStop}%
\bibitem [{\citenamefont {Braunstein}\ \emph {et~al.}(2001)\citenamefont
  {Braunstein}, \citenamefont {Fuchs}, \citenamefont {Kimble},\ and\
  \citenamefont {van Loock}}]{Braunstein2001quantum}%
  \BibitemOpen
  \bibfield  {author} {\bibinfo {author} {\bibfnamefont {S.~L.}\ \bibnamefont
  {Braunstein}}, \bibinfo {author} {\bibfnamefont {C.~A.}\ \bibnamefont
  {Fuchs}}, \bibinfo {author} {\bibfnamefont {H.~J.}\ \bibnamefont {Kimble}},\
  and\ \bibinfo {author} {\bibfnamefont {P.}~\bibnamefont {van Loock}},\
  }\bibfield  {title} {\bibinfo {title} {Quantum versus classical domains for
  teleportation with continuous variables},\ }\href@noop {} {\bibfield
  {journal} {\bibinfo  {journal} {Physical Review A}\ }\textbf {\bibinfo
  {volume} {64}},\ \bibinfo {pages} {022321} (\bibinfo {year}
  {2001})}\BibitemShut {NoStop}%
\bibitem [{\citenamefont {Xue}\ \emph {et~al.}(2019)\citenamefont {Xue},
  \citenamefont {Nguyen}, \citenamefont {James}, \citenamefont {Shabani},
  \citenamefont {Ugrinovskii},\ and\ \citenamefont
  {Petersen}}]{xue2019modeling}%
  \BibitemOpen
  \bibfield  {author} {\bibinfo {author} {\bibfnamefont {S.}~\bibnamefont
  {Xue}}, \bibinfo {author} {\bibfnamefont {T.}~\bibnamefont {Nguyen}},
  \bibinfo {author} {\bibfnamefont {M.~R.}\ \bibnamefont {James}}, \bibinfo
  {author} {\bibfnamefont {A.}~\bibnamefont {Shabani}}, \bibinfo {author}
  {\bibfnamefont {V.}~\bibnamefont {Ugrinovskii}},\ and\ \bibinfo {author}
  {\bibfnamefont {I.~R.}\ \bibnamefont {Petersen}},\ }\bibfield  {title}
  {\bibinfo {title} {Modeling for non-{M}arkovian quantum systems},\
  }\href@noop {} {\bibfield  {journal} {\bibinfo  {journal} {IEEE Transactions
  on Control Systems Technology}\ }\textbf {\bibinfo {volume} {28}},\ \bibinfo
  {pages} {2564} (\bibinfo {year} {2019})}\BibitemShut {NoStop}%
\bibitem [{\citenamefont {Gough}\ and\ \citenamefont
  {James}(2009)}]{gough2009series}%
  \BibitemOpen
  \bibfield  {author} {\bibinfo {author} {\bibfnamefont {J.}~\bibnamefont
  {Gough}}\ and\ \bibinfo {author} {\bibfnamefont {M.~R.}\ \bibnamefont
  {James}},\ }\bibfield  {title} {\bibinfo {title} {The series product and its
  application to quantum feedforward and feedback networks},\ }\href@noop {}
  {\bibfield  {journal} {\bibinfo  {journal} {IEEE Transactions on Automatic
  Control}\ }\textbf {\bibinfo {volume} {54}},\ \bibinfo {pages} {2530}
  (\bibinfo {year} {2009})}\BibitemShut {NoStop}%
\bibitem [{\citenamefont {Kitagawa}\ \emph {et~al.}(2006)\citenamefont
  {Kitagawa}, \citenamefont {Takeoka}, \citenamefont {Sasaki},\ and\
  \citenamefont {Chefles}}]{kitagawa2006entanglement}%
  \BibitemOpen
  \bibfield  {author} {\bibinfo {author} {\bibfnamefont {A.}~\bibnamefont
  {Kitagawa}}, \bibinfo {author} {\bibfnamefont {M.}~\bibnamefont {Takeoka}},
  \bibinfo {author} {\bibfnamefont {M.}~\bibnamefont {Sasaki}},\ and\ \bibinfo
  {author} {\bibfnamefont {A.}~\bibnamefont {Chefles}},\ }\bibfield  {title}
  {\bibinfo {title} {Entanglement evaluation of non-{G}aussian states generated
  by photon subtraction from squeezed states},\ }\href@noop {} {\bibfield
  {journal} {\bibinfo  {journal} {Physical Review A}\ }\textbf {\bibinfo
  {volume} {73}},\ \bibinfo {pages} {042310} (\bibinfo {year}
  {2006})}\BibitemShut {NoStop}%
\bibitem [{\citenamefont {Breuer}\ \emph {et~al.}(2009)\citenamefont {Breuer},
  \citenamefont {Laine},\ and\ \citenamefont {Piilo}}]{breuer2009measure}%
  \BibitemOpen
  \bibfield  {author} {\bibinfo {author} {\bibfnamefont {H.~P.}\ \bibnamefont
  {Breuer}}, \bibinfo {author} {\bibfnamefont {E.~M.}\ \bibnamefont {Laine}},\
  and\ \bibinfo {author} {\bibfnamefont {J.}~\bibnamefont {Piilo}},\ }\bibfield
   {title} {\bibinfo {title} {Measure for the degree of non-{M}arkovian
  behavior of quantum processes in open systems},\ }\href@noop {} {\bibfield
  {journal} {\bibinfo  {journal} {Physical Review Letters}\ }\textbf {\bibinfo
  {volume} {103}},\ \bibinfo {pages} {210401} (\bibinfo {year}
  {2009})}\BibitemShut {NoStop}%
\bibitem [{\citenamefont {Tubsrinuan}(2021)}]{tubsrinuan2021probing}%
  \BibitemOpen
  \bibfield  {author} {\bibinfo {author} {\bibfnamefont {N.}~\bibnamefont
  {Tubsrinuan}},\ }\bibfield  {title} {\bibinfo {title} {Probing two-level
  systems with a surface acoustic wave resonator},\ }\href@noop {} {\
  (\bibinfo {year} {2021})}\BibitemShut {NoStop}%
\bibitem [{\citenamefont {Burnett}\ \emph {et~al.}(2019)\citenamefont
  {Burnett}, \citenamefont {Bengtsson}, \citenamefont {Scigliuzzo},
  \citenamefont {Niepce}, \citenamefont {Kudra}, \citenamefont {Delsing},\ and\
  \citenamefont {Bylander}}]{burnett2019decoherence}%
  \BibitemOpen
  \bibfield  {author} {\bibinfo {author} {\bibfnamefont {J.~J.}\ \bibnamefont
  {Burnett}}, \bibinfo {author} {\bibfnamefont {A.}~\bibnamefont {Bengtsson}},
  \bibinfo {author} {\bibfnamefont {M.}~\bibnamefont {Scigliuzzo}}, \bibinfo
  {author} {\bibfnamefont {D.}~\bibnamefont {Niepce}}, \bibinfo {author}
  {\bibfnamefont {M.}~\bibnamefont {Kudra}}, \bibinfo {author} {\bibfnamefont
  {P.}~\bibnamefont {Delsing}},\ and\ \bibinfo {author} {\bibfnamefont
  {J.}~\bibnamefont {Bylander}},\ }\bibfield  {title} {\bibinfo {title}
  {Decoherence benchmarking of superconducting qubits},\ }\href@noop {}
  {\bibfield  {journal} {\bibinfo  {journal} {NPJ Quantum Information}\
  }\textbf {\bibinfo {volume} {5}},\ \bibinfo {pages} {54} (\bibinfo {year}
  {2019})}\BibitemShut {NoStop}%
\bibitem [{\citenamefont {Pullia}\ and\ \citenamefont
  {Riboldi}(2004)}]{pullia2004time}%
  \BibitemOpen
  \bibfield  {author} {\bibinfo {author} {\bibfnamefont {A.}~\bibnamefont
  {Pullia}}\ and\ \bibinfo {author} {\bibfnamefont {S.}~\bibnamefont
  {Riboldi}},\ }\bibfield  {title} {\bibinfo {title} {Time-domain simulation of
  electronic noises},\ }\href@noop {} {\bibfield  {journal} {\bibinfo
  {journal} {IEEE Transactions on Nuclear Science}\ }\textbf {\bibinfo {volume}
  {51}},\ \bibinfo {pages} {1817} (\bibinfo {year} {2004})}\BibitemShut
  {NoStop}%
\bibitem [{\citenamefont {Etxezarreta~Martinez}\ \emph
  {et~al.}(2021)\citenamefont {Etxezarreta~Martinez}, \citenamefont {Fuentes},
  \citenamefont {Crespo},\ and\ \citenamefont
  {Garcia-Frias}}]{etxezarreta2021time}%
  \BibitemOpen
  \bibfield  {author} {\bibinfo {author} {\bibfnamefont {J.}~\bibnamefont
  {Etxezarreta~Martinez}}, \bibinfo {author} {\bibfnamefont {P.}~\bibnamefont
  {Fuentes}}, \bibinfo {author} {\bibfnamefont {P.}~\bibnamefont {Crespo}},\
  and\ \bibinfo {author} {\bibfnamefont {J.}~\bibnamefont {Garcia-Frias}},\
  }\bibfield  {title} {\bibinfo {title} {Time-varying quantum channel models
  for superconducting qubits},\ }\href@noop {} {\bibfield  {journal} {\bibinfo
  {journal} {NPJ Quantum Information}\ }\textbf {\bibinfo {volume} {7}},\
  \bibinfo {pages} {115} (\bibinfo {year} {2021})}\BibitemShut {NoStop}%
\bibitem [{\citenamefont {Xue}\ \emph {et~al.}(2015)\citenamefont {Xue},
  \citenamefont {James}, \citenamefont {Shabani}, \citenamefont {Ugrinovskii},\
  and\ \citenamefont {Petersen}}]{xue2015quantum}%
  \BibitemOpen
  \bibfield  {author} {\bibinfo {author} {\bibfnamefont {S.}~\bibnamefont
  {Xue}}, \bibinfo {author} {\bibfnamefont {M.~R.}\ \bibnamefont {James}},
  \bibinfo {author} {\bibfnamefont {A.}~\bibnamefont {Shabani}}, \bibinfo
  {author} {\bibfnamefont {V.}~\bibnamefont {Ugrinovskii}},\ and\ \bibinfo
  {author} {\bibfnamefont {I.~R.}\ \bibnamefont {Petersen}},\ }\bibfield
  {title} {\bibinfo {title} {Quantum filter for a class of non-{M}arkovian
  quantum systems},\ }in\ \href@noop {} {\emph {\bibinfo {booktitle} {2015 54th
  IEEE Conference on Decision and Control (CDC)}}}\ (\bibinfo {organization}
  {IEEE},\ \bibinfo {year} {2015})\ pp.\ \bibinfo {pages}
  {7096--7100}\BibitemShut {NoStop}%
\bibitem [{\citenamefont {Fisher}\ \emph {et~al.}(1984)\citenamefont {Fisher},
  \citenamefont {Nieto},\ and\ \citenamefont {Sandberg}}]{PhysRevD.29.1107}%
  \BibitemOpen
  \bibfield  {author} {\bibinfo {author} {\bibfnamefont {R.~A.}\ \bibnamefont
  {Fisher}}, \bibinfo {author} {\bibfnamefont {M.~M.}\ \bibnamefont {Nieto}},\
  and\ \bibinfo {author} {\bibfnamefont {V.~D.}\ \bibnamefont {Sandberg}},\
  }\bibfield  {title} {\bibinfo {title} {Impossibility of naively generalizing
  squeezed coherent states},\ }\href@noop {} {\bibfield  {journal} {\bibinfo
  {journal} {Physical Review D}\ }\textbf {\bibinfo {volume} {29}},\ \bibinfo
  {pages} {1107} (\bibinfo {year} {1984})}\BibitemShut {NoStop}%
\bibitem [{\citenamefont {Agarwal}(2012)}]{agarwal2012quantum}%
  \BibitemOpen
  \bibfield  {author} {\bibinfo {author} {\bibfnamefont {G.~S.}\ \bibnamefont
  {Agarwal}},\ }\href@noop {} {\emph {\bibinfo {title} {Quantum optics}}}\
  (\bibinfo  {publisher} {Cambridge University Press},\ \bibinfo {year}
  {2012})\BibitemShut {NoStop}%
\bibitem [{\citenamefont {Ban}(2004)}]{ban2004phase}%
  \BibitemOpen
  \bibfield  {author} {\bibinfo {author} {\bibfnamefont {M.}~\bibnamefont
  {Ban}},\ }\bibfield  {title} {\bibinfo {title} {Phase-space approach to
  continuous variable quantum teleportation},\ }\href@noop {} {\bibfield
  {journal} {\bibinfo  {journal} {Physical Review A}\ }\textbf {\bibinfo
  {volume} {69}},\ \bibinfo {pages} {054304} (\bibinfo {year}
  {2004})}\BibitemShut {NoStop}%
\bibitem [{\citenamefont {Wiseman}\ and\ \citenamefont
  {Milburn}(1994)}]{wiseman1994all}%
  \BibitemOpen
  \bibfield  {author} {\bibinfo {author} {\bibfnamefont {H.~M.}\ \bibnamefont
  {Wiseman}}\ and\ \bibinfo {author} {\bibfnamefont {G.~J.}\ \bibnamefont
  {Milburn}},\ }\bibfield  {title} {\bibinfo {title} {All-optical versus
  electro-optical quantum-limited feedback},\ }\href@noop {} {\bibfield
  {journal} {\bibinfo  {journal} {Physical Review A}\ }\textbf {\bibinfo
  {volume} {49}},\ \bibinfo {pages} {4110} (\bibinfo {year}
  {1994})}\BibitemShut {NoStop}%
\bibitem [{\citenamefont {Gr{\"o}blacher}\ \emph {et~al.}(2013)\citenamefont
  {Gr{\"o}blacher}, \citenamefont {Hill}, \citenamefont {Safavi-Naeini},
  \citenamefont {Chan},\ and\ \citenamefont {Painter}}]{groblacher2013highly}%
  \BibitemOpen
  \bibfield  {author} {\bibinfo {author} {\bibfnamefont {S.}~\bibnamefont
  {Gr{\"o}blacher}}, \bibinfo {author} {\bibfnamefont {J.~T.}\ \bibnamefont
  {Hill}}, \bibinfo {author} {\bibfnamefont {A.~H.}\ \bibnamefont
  {Safavi-Naeini}}, \bibinfo {author} {\bibfnamefont {J.}~\bibnamefont
  {Chan}},\ and\ \bibinfo {author} {\bibfnamefont {O.}~\bibnamefont
  {Painter}},\ }\bibfield  {title} {\bibinfo {title} {Highly efficient coupling
  from an optical fiber to a nanoscale silicon optomechanical cavity},\
  }\href@noop {} {\bibfield  {journal} {\bibinfo  {journal} {Applied Physics
  Letters}\ }\textbf {\bibinfo {volume} {103}} (\bibinfo {year}
  {2013})}\BibitemShut {NoStop}%
\bibitem [{\citenamefont {Cygan}\ \emph {et~al.}(2013)\citenamefont {Cygan},
  \citenamefont {W{\'o}jtewicz}, \citenamefont {Domys{\l}awska}, \citenamefont
  {Mas{\l}owski}, \citenamefont {Bielska}, \citenamefont {Piwi{\'n}ski},
  \citenamefont {Stec}, \citenamefont {Trawi{\'n}ski}, \citenamefont {Ozimek},
  \citenamefont {Radzewicz}, \citenamefont {Abe}, \citenamefont {Ido},
  \citenamefont {Hodges}, \citenamefont {Lisak},\ and\ \citenamefont
  {Ciuryło}}]{cygan2013spectral}%
  \BibitemOpen
  \bibfield  {author} {\bibinfo {author} {\bibfnamefont {A.}~\bibnamefont
  {Cygan}}, \bibinfo {author} {\bibfnamefont {S.}~\bibnamefont
  {W{\'o}jtewicz}}, \bibinfo {author} {\bibfnamefont {J.}~\bibnamefont
  {Domys{\l}awska}}, \bibinfo {author} {\bibfnamefont {P.}~\bibnamefont
  {Mas{\l}owski}}, \bibinfo {author} {\bibfnamefont {K.}~\bibnamefont
  {Bielska}}, \bibinfo {author} {\bibfnamefont {M.}~\bibnamefont
  {Piwi{\'n}ski}}, \bibinfo {author} {\bibfnamefont {K.}~\bibnamefont {Stec}},
  \bibinfo {author} {\bibfnamefont {R.}~\bibnamefont {Trawi{\'n}ski}}, \bibinfo
  {author} {\bibfnamefont {F.}~\bibnamefont {Ozimek}}, \bibinfo {author}
  {\bibfnamefont {C.}~\bibnamefont {Radzewicz}}, \bibinfo {author}
  {\bibfnamefont {H.}~\bibnamefont {Abe}}, \bibinfo {author} {\bibfnamefont
  {T.}~\bibnamefont {Ido}}, \bibinfo {author} {\bibfnamefont {J.~T.}\
  \bibnamefont {Hodges}}, \bibinfo {author} {\bibfnamefont {D.}~\bibnamefont
  {Lisak}},\ and\ \bibinfo {author} {\bibfnamefont {R.}~\bibnamefont
  {Ciuryło}},\ }\bibfield  {title} {\bibinfo {title} {Spectral line-shapes
  investigation with {P}ound-{D}rever-{H}all-locked frequency-stabilized cavity
  ring-down spectroscopy},\ }\href@noop {} {\bibfield  {journal} {\bibinfo
  {journal} {The European Physical Journal Special Topics}\ }\textbf {\bibinfo
  {volume} {222}},\ \bibinfo {pages} {2119} (\bibinfo {year}
  {2013})}\BibitemShut {NoStop}%
\bibitem [{\citenamefont {Zhou}\ \emph {et~al.}(1996)\citenamefont {Zhou},
  \citenamefont {Doyle},\ and\ \citenamefont {Glover}}]{zhou1996robust}%
  \BibitemOpen
  \bibfield  {author} {\bibinfo {author} {\bibfnamefont {K.}~\bibnamefont
  {Zhou}}, \bibinfo {author} {\bibfnamefont {J.}~\bibnamefont {Doyle}},\ and\
  \bibinfo {author} {\bibfnamefont {K.}~\bibnamefont {Glover}},\ }\href
  {https://books.google.com/books?id=RPSOQgAACAAJ} {\emph {\bibinfo {title}
  {Robust and Optimal Control}}},\ Feher/Prentice Hall Digital and\ (\bibinfo
  {publisher} {Prentice Hall},\ \bibinfo {year} {1996})\BibitemShut {NoStop}%
\bibitem [{\citenamefont {Xu}\ \emph {et~al.}(2018)\citenamefont {Xu},
  \citenamefont {Hu},\ and\ \citenamefont {Liao}}]{xu2018improvement}%
  \BibitemOpen
  \bibfield  {author} {\bibinfo {author} {\bibfnamefont {X.}~\bibnamefont
  {Xu}}, \bibinfo {author} {\bibfnamefont {L.}~\bibnamefont {Hu}},\ and\
  \bibinfo {author} {\bibfnamefont {Z.}~\bibnamefont {Liao}},\ }\bibfield
  {title} {\bibinfo {title} {Improvement of entanglement via quantum
  scissors},\ }\href@noop {} {\bibfield  {journal} {\bibinfo  {journal}
  {Journal of the Optical Society of America B}\ }\textbf {\bibinfo {volume}
  {35}},\ \bibinfo {pages} {174} (\bibinfo {year} {2018})}\BibitemShut
  {NoStop}%
\bibitem [{\citenamefont {Hu}\ \emph {et~al.}(2017)\citenamefont {Hu},
  \citenamefont {Liao},\ and\ \citenamefont {Zubairy}}]{hu2017continuous}%
  \BibitemOpen
  \bibfield  {author} {\bibinfo {author} {\bibfnamefont {L.}~\bibnamefont
  {Hu}}, \bibinfo {author} {\bibfnamefont {Z.}~\bibnamefont {Liao}},\ and\
  \bibinfo {author} {\bibfnamefont {M.~S.}\ \bibnamefont {Zubairy}},\
  }\bibfield  {title} {\bibinfo {title} {Continuous-variable entanglement via
  multiphoton catalysis},\ }\href@noop {} {\bibfield  {journal} {\bibinfo
  {journal} {Physical Review A}\ }\textbf {\bibinfo {volume} {95}},\ \bibinfo
  {pages} {012310} (\bibinfo {year} {2017})}\BibitemShut {NoStop}%
\bibitem [{\citenamefont {Sun}\ \emph {et~al.}(2016)\citenamefont {Sun},
  \citenamefont {Mao}, \citenamefont {Chen}, \citenamefont {Zhang},
  \citenamefont {Jiang}, \citenamefont {Zhang}, \citenamefont {Zhang},
  \citenamefont {Miki}, \citenamefont {Yamashita}, \citenamefont {Terai},
  \citenamefont {Jiang}, \citenamefont {Chen}, \citenamefont {You},
  \citenamefont {Chen}, \citenamefont {Wang}, \citenamefont {Fan},
  \citenamefont {Zhang},\ and\ \citenamefont {Pan}}]{sun2016quantum}%
  \BibitemOpen
  \bibfield  {author} {\bibinfo {author} {\bibfnamefont {Q.~C.}\ \bibnamefont
  {Sun}}, \bibinfo {author} {\bibfnamefont {Y.~L.}\ \bibnamefont {Mao}},
  \bibinfo {author} {\bibfnamefont {S.~J.}\ \bibnamefont {Chen}}, \bibinfo
  {author} {\bibfnamefont {W.}~\bibnamefont {Zhang}}, \bibinfo {author}
  {\bibfnamefont {Y.~F.}\ \bibnamefont {Jiang}}, \bibinfo {author}
  {\bibfnamefont {Y.~B.}\ \bibnamefont {Zhang}}, \bibinfo {author}
  {\bibfnamefont {W.~J.}\ \bibnamefont {Zhang}}, \bibinfo {author}
  {\bibfnamefont {S.}~\bibnamefont {Miki}}, \bibinfo {author} {\bibfnamefont
  {T.}~\bibnamefont {Yamashita}}, \bibinfo {author} {\bibfnamefont
  {H.}~\bibnamefont {Terai}}, \bibinfo {author} {\bibfnamefont
  {X.}~\bibnamefont {Jiang}}, \bibinfo {author} {\bibfnamefont {T.~Y.}\
  \bibnamefont {Chen}}, \bibinfo {author} {\bibfnamefont {L.~X.}\ \bibnamefont
  {You}}, \bibinfo {author} {\bibfnamefont {X.~F.}\ \bibnamefont {Chen}},
  \bibinfo {author} {\bibfnamefont {Z.}~\bibnamefont {Wang}}, \bibinfo {author}
  {\bibfnamefont {J.~Y.}\ \bibnamefont {Fan}}, \bibinfo {author} {\bibfnamefont
  {Q.}~\bibnamefont {Zhang}},\ and\ \bibinfo {author} {\bibfnamefont {J.~W.}\
  \bibnamefont {Pan}},\ }\bibfield  {title} {\bibinfo {title} {Quantum
  teleportation with independent sources and prior entanglement distribution
  over a network},\ }\href@noop {} {\bibfield  {journal} {\bibinfo  {journal}
  {Nature Photonics}\ }\textbf {\bibinfo {volume} {10}},\ \bibinfo {pages}
  {671} (\bibinfo {year} {2016})}\BibitemShut {NoStop}%
\bibitem [{\citenamefont {Yonezawa}\ \emph {et~al.}(2004)\citenamefont
  {Yonezawa}, \citenamefont {Aoki},\ and\ \citenamefont
  {Furusawa}}]{yonezawa2004demonstration}%
  \BibitemOpen
  \bibfield  {author} {\bibinfo {author} {\bibfnamefont {H.}~\bibnamefont
  {Yonezawa}}, \bibinfo {author} {\bibfnamefont {T.}~\bibnamefont {Aoki}},\
  and\ \bibinfo {author} {\bibfnamefont {A.}~\bibnamefont {Furusawa}},\
  }\bibfield  {title} {\bibinfo {title} {Demonstration of a quantum
  teleportation network for continuous variables},\ }\href@noop {} {\bibfield
  {journal} {\bibinfo  {journal} {Nature}\ }\textbf {\bibinfo {volume} {431}},\
  \bibinfo {pages} {430} (\bibinfo {year} {2004})}\BibitemShut {NoStop}%
\bibitem [{\citenamefont {Xue}\ \emph {et~al.}(2021)\citenamefont {Xue},
  \citenamefont {Wu}, \citenamefont {Ma}, \citenamefont {Li},\ and\
  \citenamefont {Jiang}}]{xue2021gradient}%
  \BibitemOpen
  \bibfield  {author} {\bibinfo {author} {\bibfnamefont {S.}~\bibnamefont
  {Xue}}, \bibinfo {author} {\bibfnamefont {R.}~\bibnamefont {Wu}}, \bibinfo
  {author} {\bibfnamefont {S.}~\bibnamefont {Ma}}, \bibinfo {author}
  {\bibfnamefont {D.}~\bibnamefont {Li}},\ and\ \bibinfo {author}
  {\bibfnamefont {M.}~\bibnamefont {Jiang}},\ }\bibfield  {title} {\bibinfo
  {title} {Gradient algorithm for {H}amiltonian identification of open quantum
  systems},\ }\href@noop {} {\bibfield  {journal} {\bibinfo  {journal}
  {Physical Review A}\ }\textbf {\bibinfo {volume} {103}},\ \bibinfo {pages}
  {022604} (\bibinfo {year} {2021})}\BibitemShut {NoStop}%
\end{thebibliography}
\providecommand{\noopsort}[1]{}\providecommand{\singleletter}[1]{#1}%

\end{document}